\documentclass{aa}  

\usepackage{comment}
\usepackage{graphicx}
\usepackage{txfonts}
\usepackage[dvipsnames]{xcolor}
\usepackage{orcidlink}
\usepackage{colortbl}
\usepackage{txfonts}
\usepackage{hyperref}  
\hypersetup{colorlinks=true,linkcolor=[rgb]{1.,0.2,0.2},citecolor=[rgb]{0.1,0.4,1.},filecolor=[rgb]{0.7,0.2,0.2},urlcolor=[rgb]{0.7,0.2,0.2}}

\usepackage{color}
\definecolor{blue}{rgb}{0., 0., 1}

\begin{document}

   \title{Radio selection of heavily obscured AGN in the J1030 field: unraveling a missing Compton-thick population}

   \subtitle{}

   \author{Giovanni Mazzolari
          \inst{1,2,3}$^{\orcidlink{0009-0005-7383-6655}}$\fnmsep
          \thanks{giovanni.mazzolari@inaf.it},
          Roberto Gilli\inst{2}$^{\orcidlink{0000-0001-8121-6177}}$,
          Marco Mignoli\inst{2}$^{\orcidlink{0000-0002-9087-2835}}$,
          Marcella Brusa\inst{1,2}$^{\orcidlink{0000-0002-5059-6848}}$,
          Isabella Prandoni \inst{4}$^{\orcidlink{0000-0001-9680-7092}}$,
          Fabio Vito \inst{2}$^{\orcidlink{0000-0003-0680-9305}}$,
          Ivan Delvecchio \inst{2}$^{\orcidlink{0000-0001-8706-2252}}$,
          Giorgio Lanzuisi \inst{2}$^{\orcidlink{0000-0001-9094-0984}}$,
          Alessandro Peca \inst{5,6}$^{\orcidlink{0000-0003-2196-3298}}$,
          Andrea Comastri \inst{2}$^{\orcidlink{0000-0003-3451-9970}}$,
          Stefano Marchesi \inst{1,2,7}$^{\orcidlink{0000-0001-5544-0749}}$,
          Marco Chiaberge \inst{10,11}$^{\orcidlink{0000-0003-1564-3802}}$,
          Marisa Brienza \inst{4}$^{\orcidlink{0000-0003-4120-9970}}$,
          Cristian Vignali \inst{2}$^{\orcidlink{0000-0002-8853-9611}}$,
          Matilde Signorini \inst{8,15}$^{\orcidlink{0000-0002-8177-6905}}$,
          Quirino D'Amato \inst{6}$^{\orcidlink{0000-0002-9948-0897}}$,
          Fabrizio Gentile \inst{16}$^{\orcidlink{0000-0002-8008-9871}}$,
          Kazushi Iwasawa\inst{9,10}$^{\orcidlink{0000-0002-4923-3281}}$,
          Colin Norman \inst{12,13}$^{\orcidlink{0000-0002-5222-5717}}$,
          Alberto Traina \inst{2}$^{\orcidlink{0000-0003-1006-924X}}$,
          Federica Loiacono \inst{2}$^{\orcidlink{0000-0002-8858-6784}}$,
          Pietro Baldini\inst{3}$^{\orcidlink{0000-0003-1006-924X}}$,
          Marianna Annunziatella \inst{14}$^{\orcidlink{0000-0002-8053-8040}}$,
           Roberto Decarli \inst{2}$^{\orcidlink{0000-0002-2662-8803}}$.
          }

\authorrunning{G. Mazzolari et al.}
   \institute{  
        Dipartimento di Fisica e Astronomia, Università di Bologna, Via Gobetti 93/2, I-40129 Bologna, Italy
        \and 
        INAF – Osservatorio di Astrofisica e Scienza dello Spazio di Bologna, Via Gobetti 93/3, I-40129 Bologna, Italy
        \and
        Max-Planck-Institut für extraterrestrische Physik (MPE), Gießenbachstraße 1, 85748 Garching, Germany
        \and
            INAF – Istituto di Radioastronomia, Via Gobetti 101, I-40129 Bologna, Italy   
        \and
            Eureka Scientific, 2452 Delmer Street, Suite 100, Oakland, CA 94602-3017, USA
        \and 
            Department of Physics, Yale University, P.O. Box 208120, New Haven, CT 06520, USA
        \and 
            Department of Physics and Astronomy, Clemson University,  Kinard Lab of Physics, Clemson, SC 29634, USA
        \and
             INAF - Osservatorio Astrofisico di Arcetri, Largo Enrico Fermi 5, I-50125 Firenze, Italy
        \and
            Institut de Ciències del Cosmos (ICCUB), Universitat de Barcelona (IEEC-UB), Martí i Franquès, 1, 08028, Barcelona, Spain
        \and 
            ICREA, Pg. Luís Companys 23, 08010 Barcelona, Spain
        \and 
            Space Telescope Science Institute for the European Space Agency (ESA), ESA Office, 3700 San Martin Drive, Baltimore, MD, USA
        \and
            The William H. Miller III Department of Physics \& Astronomy, Johns Hopkins University, Baltimore, MD, USA
        \and
            Space Telescope Science Institute, 3700 San Martin Drive, Baltimore, MD 21218, USA
        \and
            Centro de Astrobiología, (CAB, CSIC-INTA), Carretera de Ajalvir km 4, E-28850 Torrejón de Ardoz, Madrid, Spain
        \and 
            European Space Agency (ESA), European Space Research and Technology Centre (ESTEC), Keplerlaan 1, 2201 AZ Noordwijk, The Netherlands
        \and CEA, IRFU, DAp, AIM, Université Paris-Saclay, Université Paris Cité, Sorbonne Paris Cité, CNRS, 91191 Gif-sur-Yvette, France
             }

   \date{}

 
  \abstract
   {Models of supermassive black holes (SMBHs) and galaxy coevolution, simulations, and recent JWST observations suggest that the population of heavily obscured, Compton-thick (CTK), active galactic nuclei (AGN) at high-redshift might be underestimated by X-ray surveys. To retrieve a complete census of SMBHs it is therefore necessary to identify new and complementary methods to select these sources, for example by exploiting the radio band, being radio waves almost unaffected by obscuration.}
   {We aim to test the effectiveness of radio selection to discover heavily obscured AGNs, particularly at high-z, and, in turn, measure their abundance for the first time from a radio perspective.}
   {We consider the radio sources detected in the J1030 field, which is one of the fields with the deepest combination of 1.4 GHz radio and X-ray observations publicly available. We defined a radio excess parameter as the ratio between the star formation rate (SFR) that would correspond to the observed radio luminosity and the one directly derived from the spectral energy distribution (SED) fitting, $\rm REX=SFR_{1.4 GHz}/SFR^{corr}_{SED}$. We then select as radio excess AGN those sources with $\rm REX>8.5$, corresponding to a $3\sigma$ excess above the median value ($\rm REX\sim1$). In this way, we find 145 radio-excess sources falling into the \textit{Chandra} X-ray image footprint but without X-ray detection. }
   {From the deep X-ray upper limits, we estimated a lower limit to the obscuration of each radio-excess AGN, finding on average $\log (N_H/\rm{cm^{-2}})>23.7$. A CTK AGN scenario is also supported by the results of the X-ray stacking analysis performed on sources at $z>1.5$, which revealed X-ray luminosities and hardness ratios compatible with very highly obscured AGN. Finally, we computed the number density of these radio-selected CTK AGN. While at $z\sim 2$ the radio number density agrees well with the CTK AGN predictions of different population synthesis models, at $z\sim3$ the radio selection returns a CTK AGN number density $\sim 2-3$ times larger than what is predicted by the CXB models and X-ray observations. This result supports the effectiveness of radio emission in selecting the most obscured sources, unraveling a population of AGN potentially missed by X-rays surveys at $z>3$, paving the way to a synergistic use of the future radio and X-ray facilities such as the Square Kilometer Array Observatory (\textit{SKAO}), \textit{NewAthena} and the Advance X-ray Imager Satellite (\textit{AXIS}).}
   {}


   \maketitle
%

\section{Introduction}
Models describing the formation and evolution of Supermassive black holes (SMBH) and their host galaxy suggest that obscured AGN represent a key phase during which the growth of the central SMBH is maximum and the SMBH-host galaxy scaling relations observed in the local universe are set \citep{hickox18,Hopkins08, Ferrarese05}. Traditional wide-area surveys, like Pan-STARRS, SDSS, and SHELLQs \citep{banados16,matsuoka19,fan06}, allowed the discovery of a large number of active galactic nuclei (AGN) at $z>4-5$, the large majority of them being unobscured since they were selected based on their rest-UV or optical emission, which is largely absorbed in obscured environments. This implies that previous studies are probably missing the large majority of the obscured AGN population due to selection bias. Having a proper understanding of the actual AGN demography, in particular at high redshift, is crucial to test not only the predictions of coevolutional models but also theoretical models of seed black hole formation, given that different seed black holes populations and formation efficiency translate into a different AGN census \citep{Volonteri16}.\\
Deep X-ray surveys have been traditionally employed to trace the population of obscured AGN across cosmic time, thanks to the low galaxy contamination in the X-rays at the typical AGN luminosities \citep[$L_{X}>10^{42}$erg s$^{-1}$, see][]{Lehmer16,Ranalli03}. Moreover, X-ray emission is able to penetrate through dense environments, being only marginally absorbed, at least up to Compton-thick (CTK) hydrogen column densities (defined based on the Thompson cross section $\sigma_T=6.65\times 10^{-25}$ as $N_H \sim 1/ \sigma_T\sim 1.5\times 10^{24} cm^{-2}$) and above which even the X-ray emission is  drastically suppressed. More precisely, at  $\rm \log (N_H / cm^{-2}) \sim 24$ the emission in the rest frame 0.5-10keV band is suppressed by a factor of $\sim$10-100 \citep{Brandt15,gilli07}, and most of the remaining emission in this band comes from scattered or reflected components \citep{Comastri04, Ricci15, Ricci17, Marchesi19,marchesi19b}.  \\
The results coming from X-ray surveys showed that the fraction of AGN obscured by gas column densities larger than $\rm \log (N_H / cm^{-2}) > 23$ increases up to $\sim 80\%$ at $z\sim 4$ \citep{vito14,  vito18}, with different works finding similar trends but different obscured AGN fraction, mainly depending on the methodology adopted \citep[e.g. spectroscopy vs X-ray photometry, photo-z quality, etc., see also][]{ueda14, aird15, buchner15, marchesi16b,ananna19, Signorini23, peca22}. This result is further supported by analytical models studying the evolution of the physical properties of the inter-stellar medium (ISM) around AGN and also by numerical simulations \citep{ni20,lapi20,gilli22}.\\
However, comparing the black hole accretion rate density (BHARD) derived from deep X-ray survey with those derived from theoretical models, different works noted that they are in good agreement up to $z < 3$ \citep{vito18,Peca23,Pouliasis24}, whereas at larger redshifts the models seem to overpredict the BHARD derived from X-ray surveys \citep{sijacki15,Volonteri16,shankar14}. While this discrepancy might be due to an overprediction of low-accreting SMBH in simulations \citep{Volonteri16,vito16,vito18,Habouzit21,Habouzit22,Haidar22}, or to a high dark matter halo mass selection bias of X-rays \citep{Pouliasis24}, one of the most widely proposed solutions to solve this tension is the existence of a highly obscured AGN population at high redshifts, which has been missed by X-ray surveys \citep{barchiesi21,Lyu23}. In the coevolution scenario, a larger BHARD at high-z is sought to match the high-z trend of the (rescaled) star formation rate density (SFRD), which is found to be much flatter than the X-ray BHARD at $z>3$ \citep{Madau14, Traina24a, Gentile25} \\
This possibility is further supported by some recent results obtained with James Webb Space Telescope (JWST) data. \cite{yang23} and also \cite{Hsieh25}, taking advantage of the JWST-MIRI photometry of the Cosmic Evolution Early Release Science Survey \citep[CEERS;][]{Finkelstein22}, investigated the AGN population using spectral energy distribution (SED) modeling, and found a black-hole accretion rate density (BHARD) at $z>3$ $\sim 0.5$ dex higher than what was expected from previous X-ray AGN studies and more in line with coevolutional models. Also \cite{Lyu23} performing a similar analysis on the JWST/MIRI data of the Systematic Mid-infrared Instrument Legacy Extragalactic Survey (SMILES) survey selected a remarkable fraction of AGN among the MIRI detected sources and found a statistically significant increase of the obscured AGN fraction with both AGN luminosity and redshift.  Results from \cite{Akins24} and \cite{Inayoshi24_obscuredLRD} on the selection and number density of $5\lesssim z\lesssim9$ JWST discovered AGN as the so called 'Little Red Dots' \citep[LRDs,][]{Matthee23,Greene23, Hviding22} found a BHARD 10 to 100 times larger compared to what expected from X-rays selected AGN and indeed almost none of them is X-ray detected \citep{Maiolino24_X,Yue24,Ananna24,Mazzolari25}. These results are exceeding also some of the theoretical prediction and should be taken with caution given the high uncertainties on their bolometric luminosities and black hole masses \citep{Greene25, Rusakov25}. \\
The primary reason AGN are missed in high-redshift X-ray surveys is heavy gas and dust obscuration, although other factors can also contribute. In principle, higher redshifts sample higher rest-frame X-ray energies that are less affected by obscuration effect. Nevertheless, obscurations around $\rm \log (N_H / cm^{-2}) \sim 25$ can depress the observed 2-10 keV flux by a factor of 10 even at $z\sim 3$ \citep{Hasinger08,Marchesi19}. Large obscurations can be achieved more easily at high-z, since the evolution with $z$ of the host galaxy sizes (and therefore gas density) can naturally lead to $\rm \log (N_H / cm^{-2}) \sim 24$ at $z\sim 6$ \citep{gilli22}, and this contribution is added on top of the ones coming from the evolution with $z$ of the circumnuclear gas \citep{Maiolino24_X,hickox18} and possibly of the torus covering factor \citep{lafranca05,Netzer15, Toba21}. Additionally, the many Type I AGN discovered by JWST at $z>4$ and without any X-ray detection \citep[e.g.]{Comastri25} down the the faintest fluxes of the Chandra Deep Field South and North \citep{luo17, xue16}, are posing several question on whether their X-ray weakness should be attributed to larger broad-line region covering factor or to an intrinsic weakeness due to super Eddingthon accretion \citep{Maiolino24_X,King24,Madau24}. 
It is therefore necessary to look at this population of heavily obscured AGN using multiple and complementary approaches to possibly retrieve a complete AGN census, in particular at early times. In this work, we investigate the possibility of selecting CTK AGN based on their radio emission.\\

In \cite{Mazzolari24}, we developed an analytical model to predict the number of AGN at different levels of obscuration detectable over a radio image of a given area and flux limit. By taking the radio and X-ray images of some known extragalactic deep fields, we showed that, on average, the surface density of CTK AGN detectable in the existing radio images is an order of magnitude larger than in the correspondent X-ray images. This offers the remarkable, yet unexplored, opportunity to probe, with unprecedented statistics, the most heavily obscured AGN population through radio selection.\\
The AGN spectral energy distribution (SED) is generally dominated in the radio regime by optically thin synchrotron emission that has the great advantage of being largely unaffected by obscuration \citep{padovani15,Mazzolari24,hickox18}. Historical radio surveys were mostly sensitive to the powerful ($\rm L_{1.4\, GHz}>10^{25} \ W \, Hz^{-1}$)  population of the so-called radio-loud (RL) AGN \citep{white97,becker95,becker01}, for which a significant fraction of the power produced by the accretion processes is released in kinetic form through relativistic jets that may expand up to the Mpc scales. However, new-generation radio surveys \citep{heywood20,alberts20,vandervlugt20,hale23}, reaching the $\sim \rm \mu Jy$ depth, are deep enough to detect also the radio emission coming from so-called radio-quiet (RQ) AGN, which are radiatively efficient AGN with much weaker radio emission, typically confined on (sub)kiloparsec scales. The population of RQ AGN are largely dominant over the RL one, with RL AGN being only $\sim 10\%$ of the global radio AGN population without significant dependence on redshift \citep{liu17}. Different mechanisms have been proposed in the literature to explain the origin of the radio emission in RQ AGN \citep[see][ for a review]{panessa19}, such as a coronal origin, shocks from gas outflowing from the innermost part of the AGN, and collimated young (or aborted) radio jets \citep{Chen23,Chen25}. \\
However, synchrotron radio emission is also produced by star formation (SF) processes, in particular by supernovae explosions, and in AGN host galaxies they can dilute the AGN-related radio emission. Radio excess selection techniques have been developed to distinguish between accretion, and SF-dominated radio sources \citep{delvecchio17,delvecchio21, smolcic17b,bonzini13, Zhang25,zhu23,Wang24_REX,Eberhard25}, and have now been tested to be effective out to $z>4$ \citep{Wang24_REX}. These techniques rely on the fact that AGN (even when RQ) are expected to have a radio emission in excess of what would be observed in non-active galaxies with the same stellar mass and at the same redshift because of the additional nuclear contribution. Typical parameters used to measure the AGN radio excess are, for example, those relying on the Far-Infrared Radio Correlation (FIRC) observed for SFG, like the ratio between the total infrared (IR) and the 1.4 GHz luminosity, $q_{\rm TIR}=\log(L_{\rm TIR})/\log(L_{\rm 1.4 GHz})$ \citep{novak17,delvecchio21}, or the ratio between the flux at $24\mu$m and the one at 1.4 GHz, $q_{24}=S_{24}/S_{1.4}$ \citep{bonzini13,padovani15}, or the ratio between the SFR computed from the radio luminosity and the one derived from the optical and/or IR photometry \citep{ArangoToro23,Best23,Zhang25}. These radio excess selections are able to reliably identify not only the most powerful radio AGN but also a consistent fraction of the RQ AGN population provided that they possess a dominant AGN-driven radio component, independently of which exact mechanism is responsible for it \citep{whittam22,Lyu22,Zhang25}. Interestingly, \cite{Wang24_REX} recently found that the fraction of radio-AGN increases toward higher redshift in both star forming and quiescent host galaxies.

\noindent In this work, we investigate the population of radio excess AGN in the field around the quasar SDSS J1030+0524 (hereafter J1030 field), characterized by one of the deepest combinations of radio and X-ray observations, and focusing in particular on those sources that are expected to be high-z CTK AGN. In this way, we will test the efficiency of deep radio surveys in selecting the most obscured AGN, and, in turn, we will measure their abundance for the first time from a radio perspective.\\ 
In Sect.\ref{sec:data} we present the photometric data available on the J1030 field, ranging from radio to X-rays, as well as the photometric redshifts of the radio sources. In Sect.\ref{sec:methods}, we describe the methods used to compute the main physical quantities needed for the selection of radio-excess AGN, and we define our radio-excess parameter. In Sect.~\ref{sec:Results} we present the sample of radio-excess selected AGN candidates and we compute the expected obscuration value for those that are not X-ray detected in the deep \textit{Chandra} image. In Sect.~\ref{sec:discussion} we discuss the possible contaminants of our AGN selection and we perform a detailed X-ray stacking analysis to prove both the AGN nature of the radio-excess sources and also the high-level of obscuration of those that are not X-ray detected. Finally, we compute the radio-selected CTK AGN number density, and we compare it with results coming from X-ray observations and models. We summarize the results of our work in Section \ref{sec:conclusions}.

We assume a flat $\Lambda$CDM universe with $H_{0}=70\ \rm{km s^{-1} Mpc^{-1}}$, $\Omega_{m}=0.3, \Omega_{\Lambda}=0.7$.
We assume AGN radio spectra of the form $S_{\nu}\propto \nu^{-\alpha}$, with $\alpha=0.7$, which is the typical spectral slope considered for extragalactic synchrotron emission \citep{smolcic17a,novak17}. We refer to $\rm \nu L_{\nu}$ (with $\rm \nu=1.4 GHz$), in units of $\rm erg \, s^{-1}$ as $\rm L_{1.4\rm GHz}$.

\section{Data} \label{sec:data}

\subsection{J1030 multiband catalog} \label{sec:J1030cat}
\begin{figure}
    \centering
   \includegraphics[width=1\linewidth]{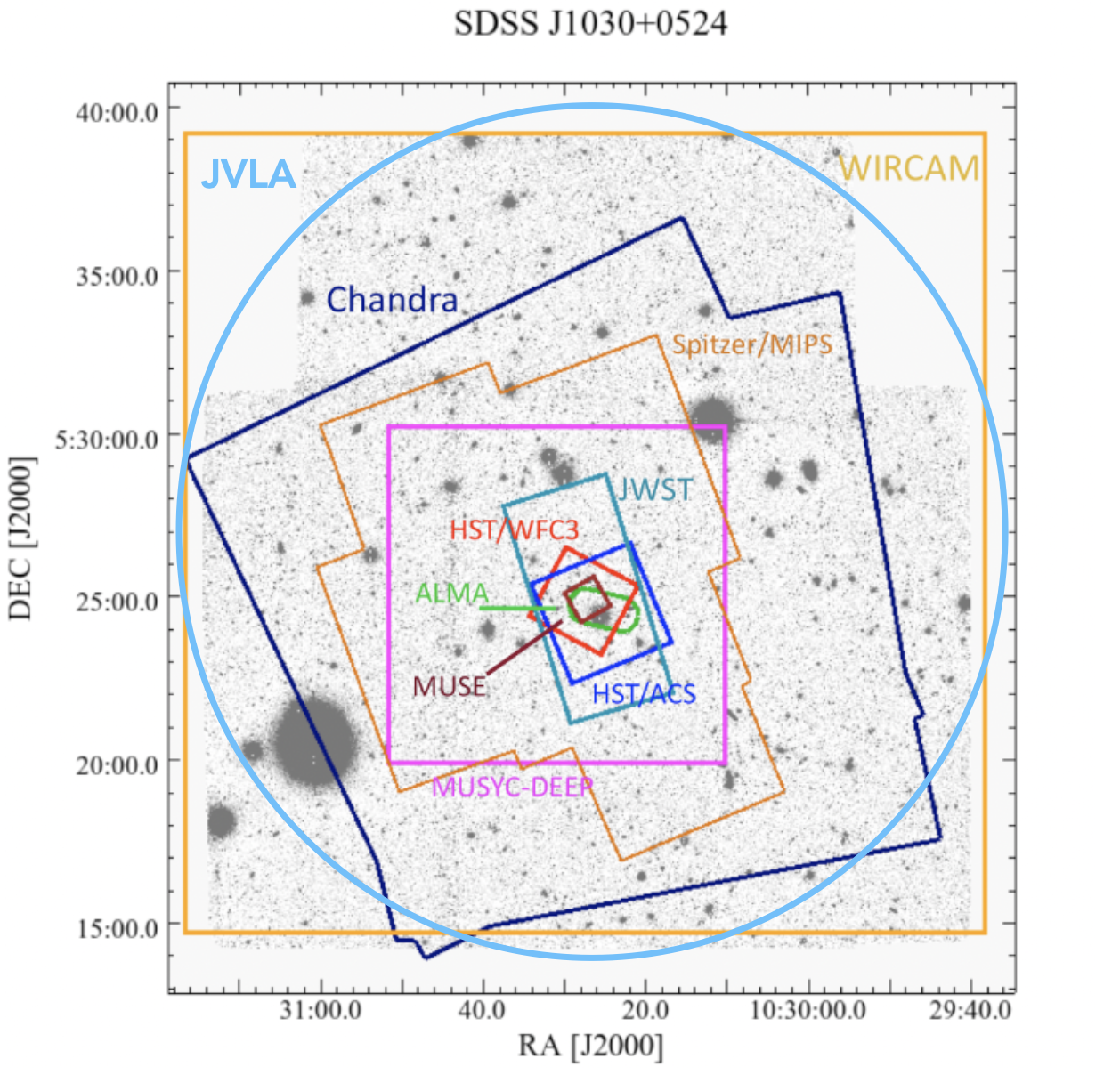}
      \caption{LBT/LBC z-band image of the field. The coverage of different surveys/instruments is shown. The entire field is also covered by XMM, MUSYC wide, and Spitzer/IRAC.}
         \label{fig:J1030}
\end{figure}

The $\sim 23' \times 23'$ arcmin$^2$ region around the luminous quasar SDSS J1030+0524 (z=6.31) has been imaged at both optical (LBT/LBC in the g, r, i, z bands) and near-IR (CFHT/WIRCAM Y, J, Ks) bands. The field is part of the MUSYC survey \citep{Gawiser06}, it is entirely covered by a Spitzer IRAC mosaic \citep{annunziatella18}, and in the mm domain has been observed by AzTEC \citep{Zeballos18}. The inner square arcmin around the QSO has been observed by HST-ACS \citep{stiavelli05} and HST-WCF3, VLT-MUSE \citep{mignoli20}, and ALMA \citep{damato20}. The J1030 multiband photometric catalog, contains $\sim 15000$ Ks-band selected sources with associated aperture photometry in 11 bands, from MUSYC/U band ($\sim 350 \rm nm$) to IRAC/CH2 ($\sim 4500 \rm nm $). Details on the observations and photometry extraction will be described in Mignoli et al. in prep.\\
The central region of the J1030 field has also been observed by a JWST NIRCam mosaic (imaging and slitless spectroscopy) as part of the EIGER program \citep{Kashino23}: these data will be the subject of future works.\\
The central $17'\times17'$ arcmin of the J1030 field has also been observed by \textit{Chandra} for $\sim 500$ks \citep{nanni20}, producing one of the deepest extragalactic X-ray fields to date reaching flux limits of $\sim 3, 0.6, 2\times10^{-17}$ erg cm$^{-2}$ s$^{-1}$ in the 0.5-7 keV (full), 0.5-2 keV (soft) and 2-7 keV (hard) bands, respectively. The X-ray catalog contains 256 single X-ray sources, and their spectroscopic and photometric redshift and X-ray spectral analysis have already been presented in \cite{marchesi21, Marchesi23} and \cite{Signorini23}, respectively. In Fig.\ref{fig:J1030} we show the J1030 field covered by the main multiband observations.

\subsection{J1030 radio data}\label{sec:J1030rad}
The deep radio observation of the J1030 field and its related catalog, which covers a circular region of $\sim25$ arcmin of radius, were presented and described in detail in \cite{damato22}. The $\sim 30$hrs of JVLA observation at 1.4 GHz allowed achieving a depth of $\sim2$ $\mu$Jy beam$^{-1}$ of rms in the central region, comparable with that of the deepest extragalactic radio surveys to date \citep{whittam22,owen18,alberts20}, with an angular resolution of $\sim 1^{\prime \prime}$.
The radio catalog contains 1486 individual sources, with 1102 of them falling into  the footprint of the Ks-band selected catalog, that contains the photometric information of all the bands presented in Section~\ref{sec:J1030cat} (we will refer to this catalog as the multiband catalog hereafter). Using the counterpart matching code \texttt{NWAY} \citep{Salvato18}, we searched for their counterparts among the multiband catalog sources, using a matching radius of $1.5"$. The code returns two probabilities that can be used to determine the correct counterpart. One is the probability called \texttt{p$_{any}$}, which represents the probability that the source from the first catalog could have a counterpart among the sources of the second catalog, while the second is $p_i$, the probability of each possible counterpart association. NWAY safe matches have both $p_{any},p_i>0.5$ as it was tested in \cite{Salvato18} and applied in \cite{Salvato25}. We visually inspected all the associations with $p_{any}<0.5$ or $p_i<0.5$ to check any possible false match. The matching procedure returned 1003 radio sources with a counterpart in the multiband catalog (91\% of the radio sources in the multiband catalog footprint), while 99 sources are without a K-band counterpart.  We visually inspected these 99 sources: 45 are visible only in bands redder than the Ks band (Sapori et al. in prep.), 14 are non detected in the Ks-band but have a counterpart only in bluer bands, 18 are blended sources and the remaining 22 are without a counterpart in any of the optical or NIR bands.\\
Using the same procedure, we also cross-matched the radio sources with the X-ray ones. There are 763 radio sources with a multiband counterpart in the \textit{Chandra} footprint; 101 of them are also associated with an X-ray-detected source. 

\subsection{J1030 photometric redshift}\label{sec:J1030photoz}

For the 1003 radio sources with an associated counterpart in the J1030 multiband photometric catalog, we took the redshift reported in Mignoli et al. (in prep). When available, we used the spectroscopic redshift, but most of the sources  in the multiband catalog ($\sim 90\%$) have only a photometric redshift. Among the 1003 radio sources, only 95 have a spectroscopic redshift \citep[of these, 62 are also X-ray detected sources that were the main targets of the spectroscopic follow-up campaigns presented in ][]{marchesi21}. The photometric redshifts  of the $\sim 14000$ sources of the multiband catalog were derived by combining the results from three different photometric redshift codes: \texttt{eazypy} \citep{brammer08}, \texttt{LePhare} \citep{ilbert09}, and \texttt{Hyperz} \citep{bolzonella00}. In particular, for each code we used two different sets of templates. For \texttt{eazypy} we used two different galaxy templates implementing full stellar population synthesis models and star formation histories coherent with the redshift of the sources, while for both \texttt{LePhare} and \texttt{Hyperz} we used one set of galactic templates and one set of AGN and composite templates. To validate these photometric redshifts, we used the sample of spectroscopic redshifts of the J1030 field (283 sources). However, 17\% of the sources with spectroscopic redshift are Broad-line AGN (BLAGN) and another 10\% are Narrow-line AGN (NLAGN), as a result of the follow-up observations of the X-ray sources. Our sample is therefore biased towards AGN, whose SED, especially for BLAGN, cannot be adequately reproduced by the galaxy templates. We therefore decided to exclude BLAGN (and stars) from the reference spectroscopic sample used to validate the J1030 photometric redshifts. The final spectroscopic catalog includes 219 sources. Then, to quantify the fraction of outliers $\eta$, following \cite{Hildebrandt12}, we considered those galaxies whose photo-z deviates from their spec-z by $|\Delta z|>0.15(1+z_{spec})$. We also considered the relative outlier fraction $\eta_{rel}$ defined as the fraction of sources for which $|\Delta z|>3 \sigma_{NMAD}$.\\
The values of the different photo-z quality parameters for the different sets of code-template are reported in Table~\ref{tab:zphot_quality}.
All the photometric redshift code performs similarly in terms of $\sigma_{NMAD}$ and $\eta$, except for the solutions obtained using only AGN templates that perform significantly worse on the overall population, as expected. \\ 
Since photo-z codes usually underestimate photometric redshift uncertainties \citep{Salvato19}, we applied a smoothing procedure to the Pdz of each sources derived from each code, following the example reported in \cite{Dahlen13} \footnote{the smoothing of the Pdz in correspondence to the redshift $z_j$ (i.e. $P(z_{j})$) was done such that for each redshift bin j  (and considering a linear redshift grid with step $\Delta z=0.01$)} we replace $P(z_{j})$ with this linear combination of the adjacent bins: $P'(z_{j})= 0.25 P(z_{j-1}) + 0.5 P(z_{j}) + 0.25 P(z_{j+1})$ \citep{Dahlen13}. In particular, for each code, we iterated the smoothing procedure till the fraction of $z_{spec}$ that belongs to the interval [$z_{phot}-z_{err}$, $z_{phot}+z_{err}$] is 0.68, correctly representing the 1$\sigma$ uncertainty.\\
Since the different photometric redshift solutions perform similarly in terms of $\eta $ and $\sigma_{NMAD}$, we derived the final photometric redshift solution combining the results of the codes. As demonstrated in \cite{Dahlen13}, this approach can reduce both the effect of systematic errors associated with single codes and also the scatter of the $z_{phot}$ final solution around the true redshift. For each source, we summed the full smoothed Pdz returned by the different codes and then took as best $z_{phot}$ the value corresponding to the median of the summed Pdz. Contrary to using the strict median of the different redshifts, this approach allows the computation of the full summed probability distribution function, which can be used to compute the photometric redshift errors in a consistent way. We set as upper and lower errors the values corresponding to the 16 and 84 percentiles of the Pdz distribution.
Considering only the 1003 radio sources with a multiband counterpart, they span a redshift range $0.1<z_{phot}<6.3$, with a median redshift of $z_{phot}=1.22$.

\section{Methods}\label{sec:methods}

\subsection{ Radio and SED fitting star formation rates}\label{sec:SEDfit}
Since all the sources are radio-detected, we can compute their 1.4 GHz luminosities and then the SFR as if all the radio luminosity is due to SF processes. Following \cite{novak17}, we first computed the radio luminosity using:
\begin{equation}\label{eq:Lrad}
    L_{\nu,1.4 GHz}=\frac{4\pi d_L^2 S_{1.4 GHz}}{(1+z)^{1+\alpha}} \ \ W Hz^{-1},
\end{equation}
where $L_{\nu,1.4GHz}$ is the radio luminosity density, $d_L$ is the luminosity distance and $\alpha$ the radio spectral index. Then the corresponding radio SFR (SFR$_{\rm 1.4~ GHz}$) is given by:
\begin{equation}\label{eq:SFRrad}
    \rm SFR_{1.4 ~ GHz}= f_{IMF}\times10^{-24}10^{q_{TIR}} \frac{L_{1.4 ~ GHz}}{W ~ Hz^{-1}} \ \ M_{\odot}yr^{-1},
\end{equation}
where $f_{IMF}$ depends on the assumed IMF and corresponds to $f_{IMF}=1$ for a Chabrier IMF \citep{chabrier03} and to $f_{IMF}=1.7$ for a Salpeter one. For consistency with the results from the SED fitting, we used $f_{IMF}=1$. In Eq.~\ref{eq:SFRrad}, $q_{TIR}$ represents the ratio between the total infrared luminosity and the radio luminosity at 1.4 GHz, as derived assuming the FIRC for SFG. This correlation was found to depend both on the redshift and the stellar mass ($\rm M_{\star}$) of the sources \citep{delvecchio21}, as follows:
\begin{equation}\label{eq:qtir}
    q_{TIR}=2.646\times(1+z)^{-0.023} - 0.148\times \log \biggl(\frac{M_{\star}}{10^{10}M_{\odot}}\biggr).
\end{equation}
In particular, we used in Eq.~\ref{eq:qtir} the value of $M_{\star}$ derived from the SED fitting with \texttt{CIGALE}.\\
Indeed, to infer the physical properties of radio-detected galaxies, we performed a SED fitting analysis using \texttt{CIGALE} \citep{boquien19,yang20}, considering the eleven bands in the optical and NIR from the multiband photometric catalog. In particular, we used delayed star formation history (SFH) models that can reproduce both early-type and late-type galaxies. We adopted stellar templates from \cite{bruzual03}, and a Chabrier initial mass function. We also include the nebular emission module, which is extremely important to account for the contribution of emission lines in the broad-band photometry \citep{Schaerer2012,Salvato2019}.  We assumed solar metallicities for both the stellar and gas component. For the attenuation of the stellar continuum emission, we considered the \texttt{dustatt\_modified\_CF00} module \citep{Charlot00}, which allows different attenuations for the young and old stellar populations, and we allowed for different levels of attenuation $0.01<A_{\rm v}<3.5$. We also include dust emission in the IR following the empirical templates of \cite{Dale14}. \\
We did not include the radio data point in the fit because we do not want to contaminate the SFRs from a potential AGN-related radio emission, which would lead to an erroneous larger SFR and would limit the effectiveness of the radio excess selection, as we will describe later. According to this, we also did not include in the SED fitting any AGN module, standardizing the fit of all the sources as if they were normal SFG, deferring the AGN identification to the radio-excess analysis shown in Sect.~\ref{sec:REX}. 
Only for the X-ray detected sources, which we already know being AGN according to previous works, we performed the fit also including the AGN module in \texttt{CIGALE}. In particular, we employed the \texttt{skirtor2016} module introduced in \cite{yang20},  which provides a more realistic description of the AGN properties compared to the other \texttt{CIGALE} AGN module based on \cite{fritz2006agnmodel} and is also the most widely used module to model the multiwavelength AGN emission \citep[e.g.,][]{Mountrichas22, LopezIE23, yang23}. The SED produced by this AGN module combines emissions from the accretion disk, torus, and polar dust. For the X-ray sources, we further included the X-ray photometry in the fit \citep{nanni20,marchesi21} that is modeled by the X-ray module of \texttt{CIGALE} taking into account both the emission of AGN and X-ray binaries \citep{yang20}. We found that for 7 X-ray sources (all with $\log (L_{2-10~keV}/\rm erg\ s^{-1})<41.9$), the X-ray data point (and the fit in general) could be explained without the need of an AGN component. We will, therefore, refer to the sample of X-ray detected radio AGN for the remaining 94 sources.\\
We used the SED-fitting to derive two main parameters: the stellar mass $M_{\star}$ and the extinction corrected star formation rate $SFR^{corr}_{SED}$.

\subsection{Radio Excess parameter}\label{sec:REX}
\begin{figure}[h!]
    \centering
   \includegraphics[width=\linewidth]{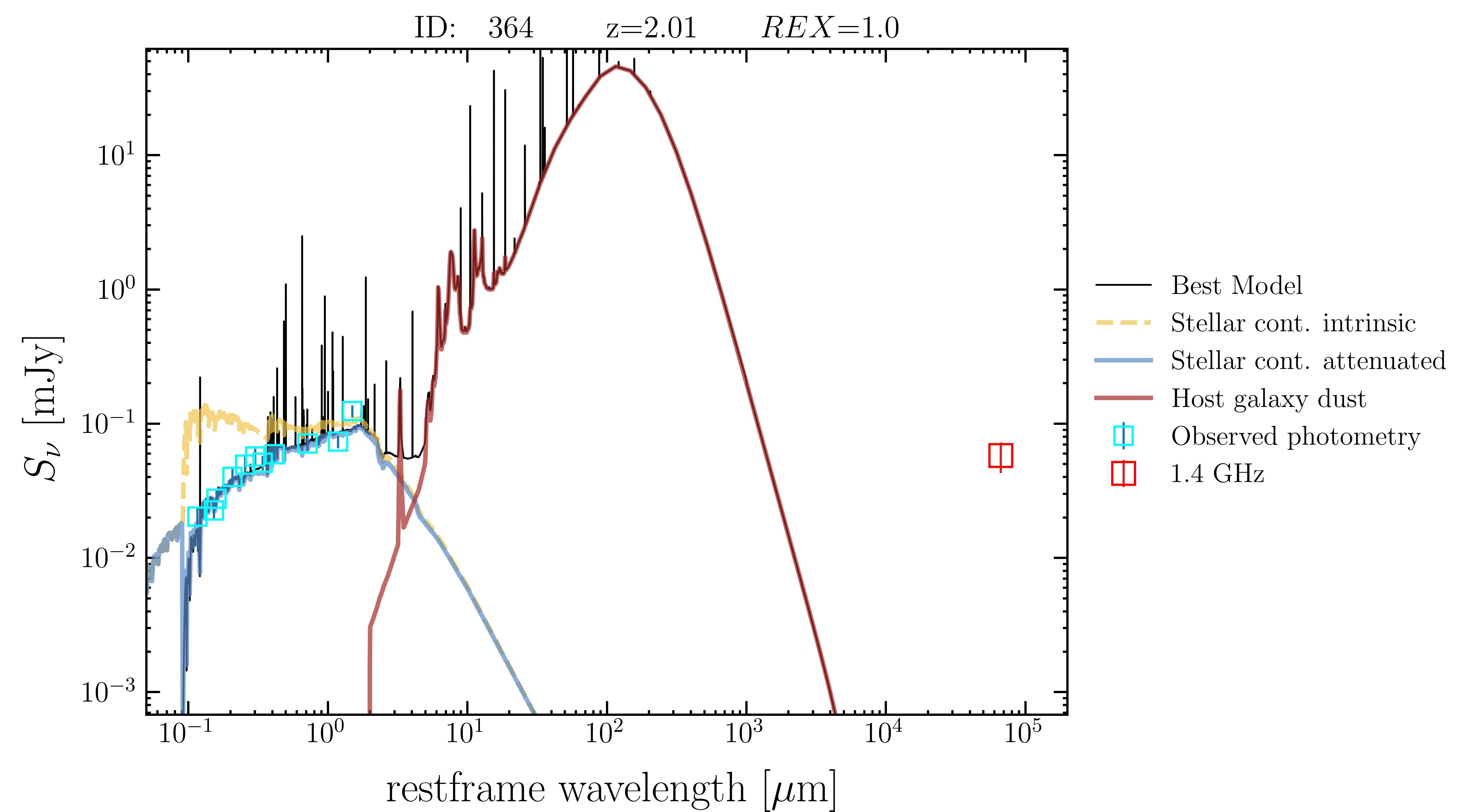}
   \includegraphics[width=\linewidth]{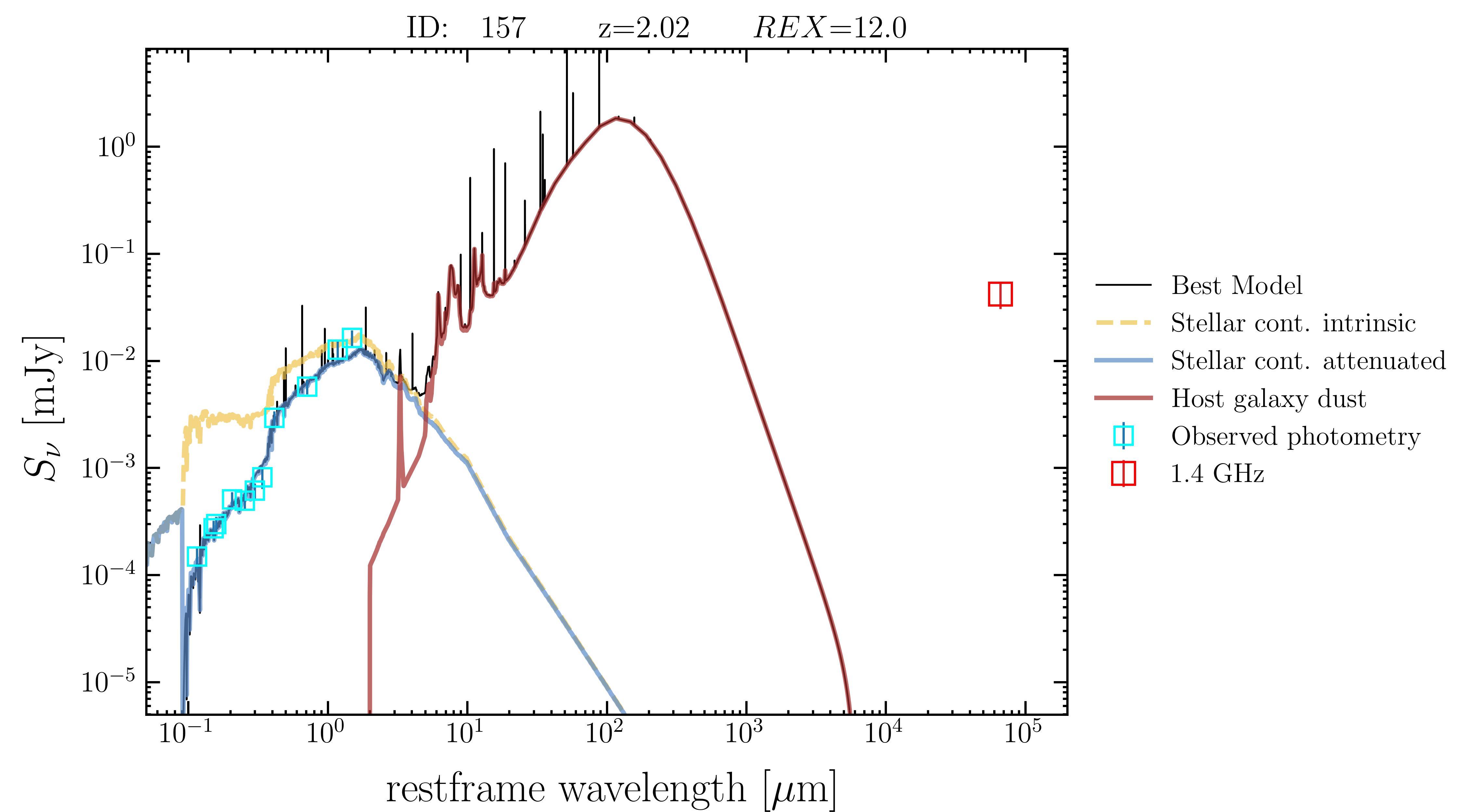}
   \includegraphics[width=\linewidth]{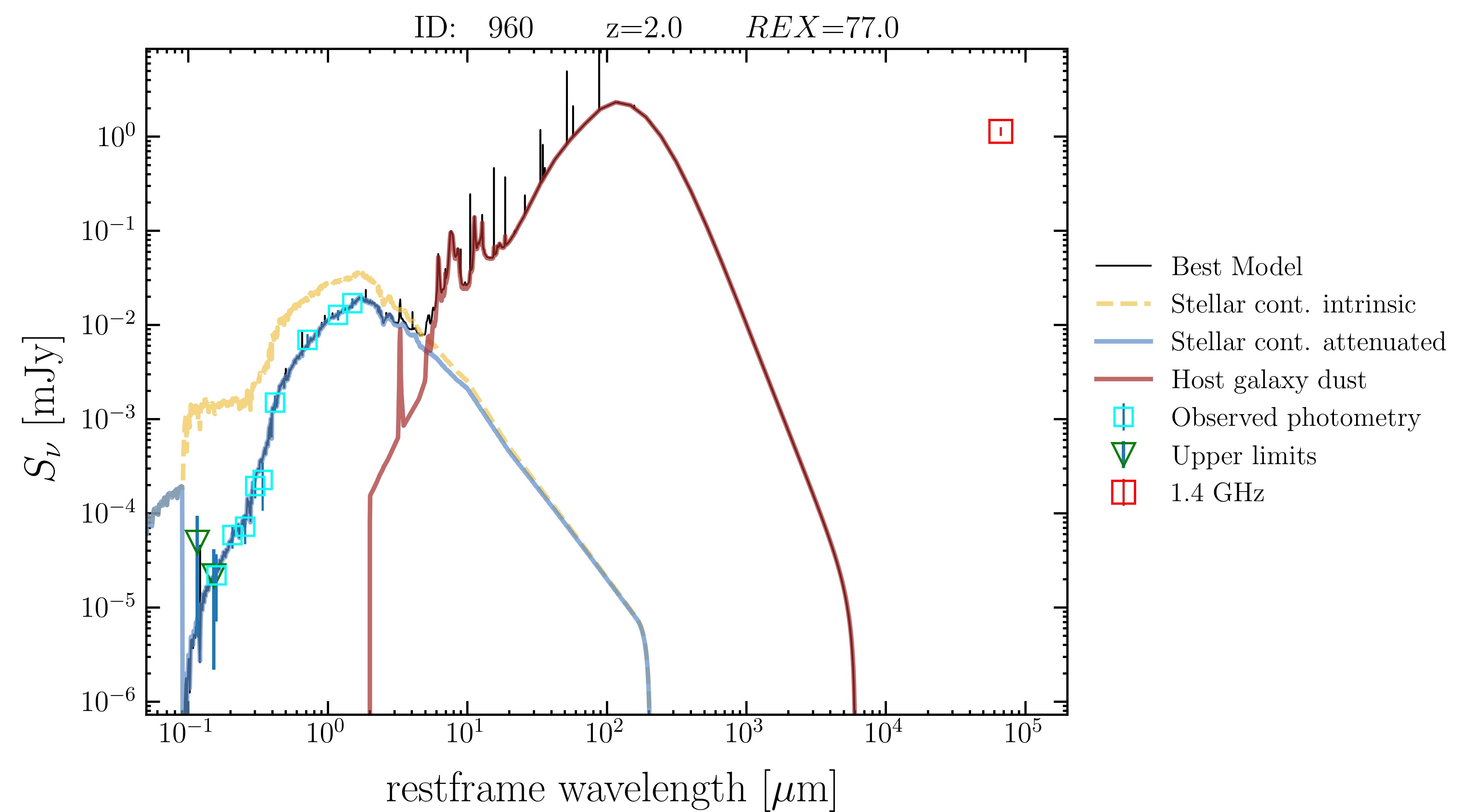}
      \caption{ SED fitting decomposition of three different sources at $z\sim2$ with different values of $REX$. The black line represents the best fit model, the yellow line the intrinsic stellar emission, the blue component is the attenuated stellar emission, while the brown component is the reprocessed dust emission (computed assuming energy balance).}
         \label{fig:SEDs}
\end{figure}

The radio-excess AGN selection techniques are based on the identification of sources whose radio emission can be justified only with an additional AGN component on top of the SF emission. We identified as radio-excess parameter the ratio between the SFR derived directly from the radio luminosity and the SFR derived from the SED fitting to the optical-NIR photometry:
\begin{equation}\label{eq:REX}
    \rm REX=\frac{SFR_{1.4 GHz}}{SFR^{corr}_{SED}}.
\end{equation}
Since the stars emitting in the rest frame UV and optical bands are the same producing the radio emission by means of supernovae explosions, we expect normal SFG (with no AGN) to be distributed around $REX\simeq 1$, with a certain scatter. On the contrary, those sources hosting an AGN should have larger values of $REX$, because of the additional radio emission connected to the nuclear processes and not to SF.  The \texttt{CIGALE} SEDs of three sources at $z\sim 2$ with different $REX$ are shown in Fig.~\ref{fig:SEDs}.

\section{Results} \label{sec:Results}

\subsection{Radio excess AGN selection}\label{sec:radioexcessAGN}
\begin{figure}
    \centering
   \includegraphics[width=\linewidth]{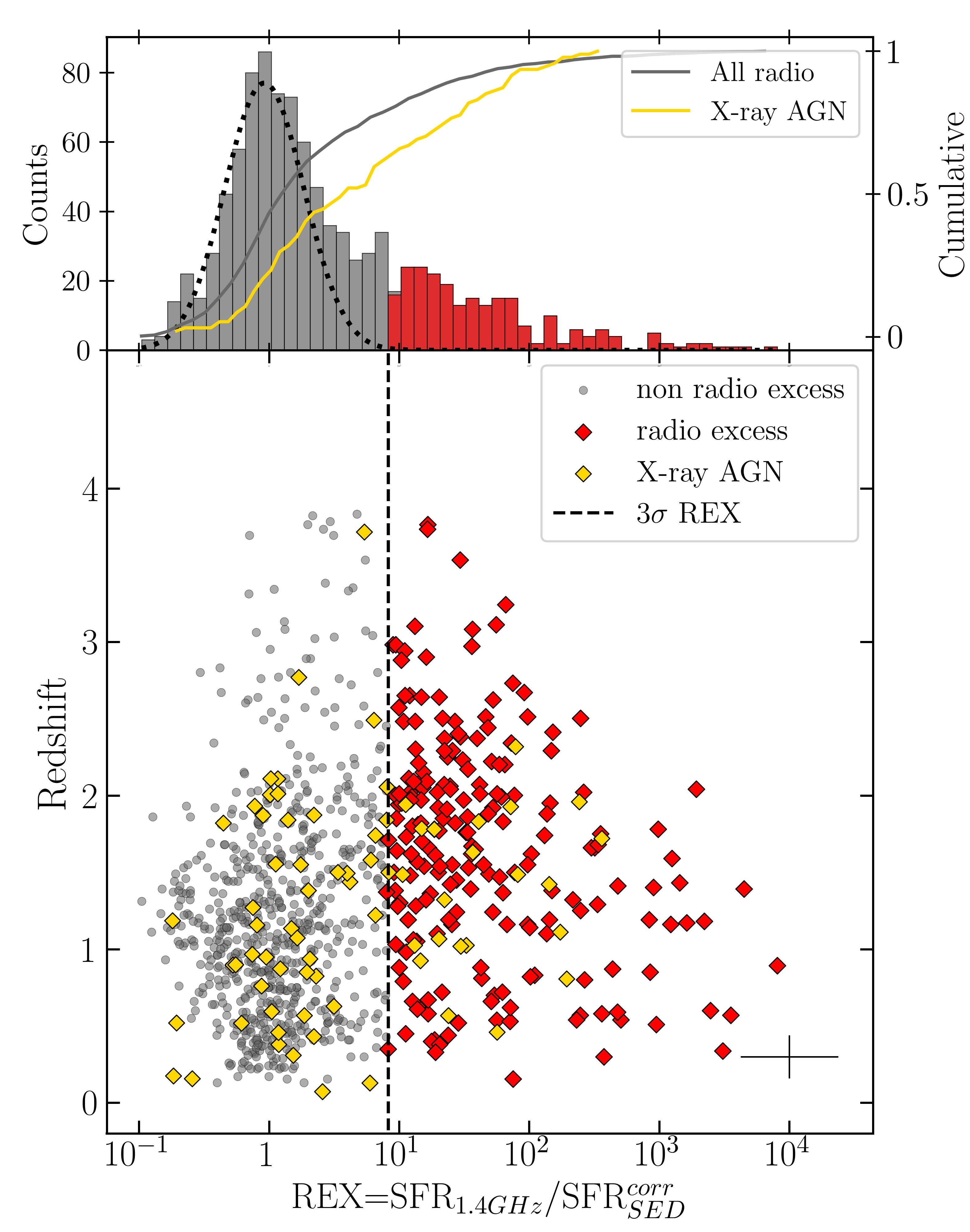}
      \caption{Redshift versus $REX$ parameter of all the radio sources in the J1030 field with a counterpart in the Ks-band selected multiwavelength catalog. The red squares represent the radio-excess sources, while gold symbols are used for X-ray detected AGN. The dashed line remarks the 3$\sigma$ deviation from the peak of the Gaussian distribution of the $REX$ parameter (upper panel), which is likely to be dominated by SFGs being $REX\simeq 1$. The histogram in the upper panel shows the distribution of the sources according to the $REX$ parameter, while the gray and gold lines are the cumulative distribution of all the radio sources and of the X-ray AGN, respectively.  In the lower right corner we report the average errors of the data points on the two axis.}
         \label{fig:REXhist}
\end{figure}

In Fig.~\ref{fig:REXhist}, we show the distribution of all the radio sources with a counterpart in the multiband photometric catalog of the J1030 field in the redshift versus $REX$ parameter plane (with $REX$ values distributed in logarithmic space). As it is possible to see from the upper panel of Fig.~\ref{fig:REXhist}, the peak of the $REX$ distribution is centered almost exactly on $REX=1$. This result supports the robustness of our SED-fitting procedure and the validity of Eq.~\ref{eq:SFRrad} for SFG. However, the $REX$ distribution is highly asymmetric, showing a tail extending up to $REX\sim 10^4$, probably dominated by AGN. Therefore, to identify AGN, we first have to model the SFG population. Considering the peak of this distribution as dominated by SFG, we mirrored the left-hand part of the log-histogram with respect to the peak and fit this mirrored distribution with a Gaussian function. The fit in the $\log REX$ space returned a $\sigma_{\log REX}=0.31$ representing the intrinsic dispersion of the SFG population around the mean $\mu_{\log REX}=-0.03$ (equivalent to $\mu_{REX}=0.92$). Therefore, we can identify as radio-excess AGN all the sources that are $3\sigma$ outliers from the peak of the SFG Gaussian, namely all those sources with:
\begin{equation}
    \rm{REX}\geq 10^{\mu_{\log REX}+3\sigma_{\log REX}}\simeq 8.5 .
\end{equation}
For these sources the radio emission that produces the excess in the observed $SFR_{1.4 ~ GHz}$ can be attributed to an AGN-related radio emission (as we will confirm in the next sections). We note that the uncertainties on the $SFR^{corr}_{SED}$ can potentially scatter objects that are not AGN into the radio excess zone. However, these uncertainties have the primary effect of broadening the Gaussian around the peak value of the distribution of $REX$. Since our radio-excess selection is not based on a fixed value but is defined in terms of $\sigma$ of the SFG distribution, the impact of SED-fitting uncertainties on the radio-excess selection is largely mitigated. Among the 1003 radio sources with a multiwavelength counterpart, we identified 233 radio-excess sources, 181 of them falling inside the \textit{Chandra} footprint and 145 without an X-ray counterpart.  \\
In Fig.~\ref{fig:REXhist}, we also show the 94 radio AGN detected in the X-rays. As shown by the cumulative distribution of these sources in $REX$ (upper panel), they are significantly shifted towards higher values of $REX$ with respect to the cumulative distribution of all the radio sources. We performed an Anderson-Darling test on the X-ray AGN and all radio sources distributions, finding a p-value$\sim 0.001$, implying that the two distributions are different at the $> 99.9\%$ confidence level. This strongly supports the hypothesis that radio-excess sources are dominated by AGN.\\
 In Appendix~\ref{app:contamination} we discuss the effect of the uncertainties on the $REX$ parameter on the radio-excess selection. Considering error propagation of all the quantities contributing to the $REX$ (i.e., SFR derived from \texttt{CIGALE}, redshift, and radio flux), we do not expect a contamination of the radio-excess selection larger than the $\sim 10\%$. This fraction is not expected to have a significant impact on the results that follow. 

\subsection{$SFR^{corr}_{SED}$ distribution of radio sources}\label{sec:SFRdistr}
\begin{figure}
    \centering
   \includegraphics[width=\linewidth]{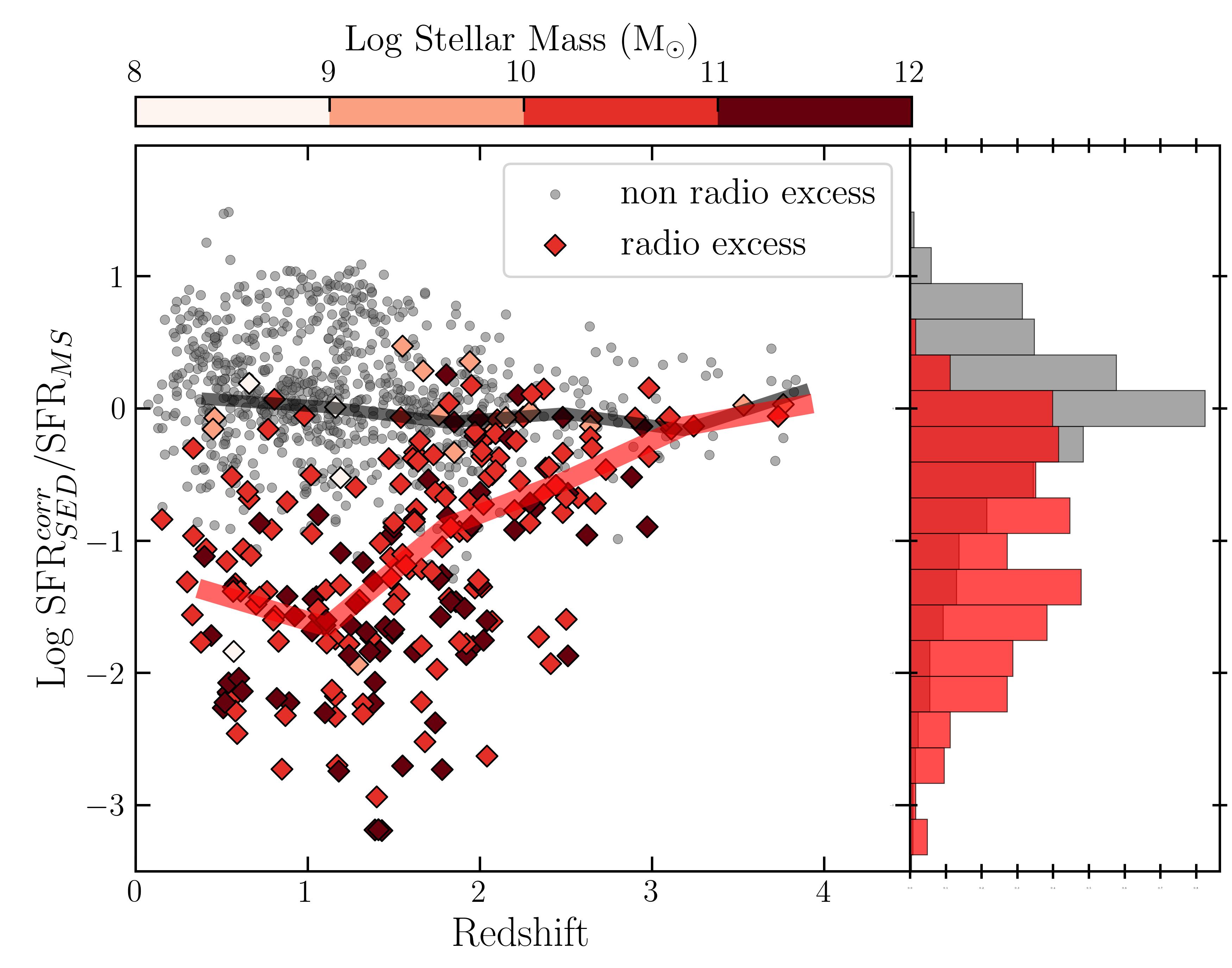}
      \caption{Distribution of the radio sources with a counterpart in the J1030 multiwavelength catalog according to their redshift and to the SFR derived from the SED-fitting normalized to the expected main sequence SFR. Radio-excess sources are also color-coded by the stellar mass returned by the SED fitting. $SFR^{corr}_{SED}$ for radio-excess sources have been computed in a second run by adding the AGN component to the fit. The red and gray lines mark the median $SFR^{corr}_{SED}$/$SFR_{MS}$ with redshift for the populations of radio-excess and non radio excess sources, respectively. On the left is plotted the density distribution according to the MS-normalized SFR (red refers to radio-excess sources).}
         \label{fig:SFR_MS}
\end{figure}
In Fig.~\ref{fig:SFR_MS}, we show the distribution of the radio sources according to their redshift and to the $SFR^{corr}_{SED}$ derived from the SED fitting normalized to their expected main-sequence (MS) $SFR$.  For this, from the redshift and the inferred $M_{\star}$, we predict the main sequence SFR ($SFR_{MS}$) using \cite{Speagle14}.
Sources are also color-coded by the stellar mass derived from the SED-fitting. For the radio-excess sources identified as AGN candidates, we now repeated the SED-fitting procedure adding the AGN component to the fit performed with \texttt{CIGALE} to obtain their actual values of $SFR^{corr}_{SED}$ and $M_{\star}$. 
From Fig.~\ref{fig:SFR_MS} we note that, in general, radio excess sources are characterized by lower $SFR^{corr}_{SED}$ with respect to what is expected from the main sequence. This is particularly true for $z<2$ and high-mass host galaxies ($\log (M_{\star}/M_{\odot})>11$), while for sources at $z> 2$ (and for lower mass host galaxies), the ratio between the two SFRs is $\sim 1$ (0 on the y axis of Fig.~\ref{fig:SFR_MS}). This distribution is expected for three main reasons. First, given our radio selection (corresponding to a SFR selection for galaxies), we are able to detect significantly low-main sequence galaxies only at lower redshifts. Second, by definition, the $REX$ parameter favors the identification of radio-excess AGN in low-SF systems. Third, different works showed that the integrated effect of the AGN feedback on the host galaxy tends to quench their SFR \citep{Piotrowska2022, Bluck2023}, determining AGN host-galaxy to generally have suppressed SFR. This effect is observed in particular at low redshifts where many radio galaxies hosting AGN show a population of old stars associated with inefficient SF processes. This distribution of the radio-excess sources is in agreement with what was found in \cite{Zhang25}: radio AGN in massive, low redshift galaxies usually tend to show low-main sequence SFR. On the contrary, at higher redshifts (as well as in low mass systems), SFR of the radio AGN host galaxies are generally aligned with the main sequence, or even above \citep{smolcic17b, Best23, Zhang25}.

\subsection{RLAGN}\label{sec:RLAGN}
The selected sample of radio excess AGN candidates certainly includes also RL AGN. To identify these sources, we took advantage of the final SED obtained including in the SED-fitting with \texttt{CIGALE} the AGN component. We computed the radio loudness parameter $R=L_{5GHz}/L_{4400\AA}$ \citep{Kellermann1989}, defined as the ratio between the radio luminosity at rest frame 5GHz and the unattenuated AGN optical luminosity at rest frame 4400$\AA$ (both computed in W$\rm Hz^{-1}$). In particular, we derived $L_{5GHz}$ by converting the radio luminosity computed in Eq.~\ref{eq:Lrad} using $\alpha=-0.7$, while we derived $L_{4400\AA}$ directly from the final SED returned by the SED fitting procedure. Following \cite{Bariuan22} we identified as radio loudness threshold the value $R=30$\footnote{This value is slightly larger than the typical threshold of $R=10$, but allows us to be more conservative in the identification of RL AGN.}. 
We also include in the RL AGN sample radio sources classified as extended (i.e. showing radio jets) in the original radio catalog of \cite{damato22} (6 sources). The selection returned 36 sources reliably classified as RL AGN included in the radio-excess sample. We will, therefore, consider the remaining sample of radio-excess AGN candidates as RQ AGN.

\subsection{Obscuration} \label{sec:obsc}
In this work, we do not want to investigate only the radio-excess AGN population, but we want to use the radio emission to uncover the population of heavily obscured AGN.\\
To do this we considered only those radio excess sources that fall in the \textit{Chandra} X-ray image footprint but that are not detected in the deep X-ray catalog. These are 145 sources. Assuming these sources to be AGN, implies that they should be heavily obscured to be non detected in the 500ks {\it Chandra} observation. To quantify the level of obscuration, we need to find the minimum level of $N_H$ that, given the redshift and the intrinsic X-ray luminosity of the source, would return a non-detection in the X-ray image at the position of the source. Therefore, we first measured the X-ray flux limits in the soft band (SB, 0.5-2 keV) at the position of the non-X-ray detected radio-excess sources. Since we know their radio luminosity, we can compute their intrinsic (i.e. absorption-corrected) X-ray luminosity using the well-known X-ray to radio luminosity relations suited for the populations of RQ and RL AGN \citep{panessa15, damato22, Wang24_FP, Bariuan22}. In particular, for the population of RQ AGN we used the $L_{1.4 ~ GHz}-L_{2-10 ~ keV}$ luminosity relation derived in \cite{damato22} considering a sample of 89 spectroscopically confirmed AGN and Early Type Galaxies (ETG) on the J1030 field extending up to $z\sim 3$. Most of the sources ($\sim 80\%$) componing the sample of \cite{damato22} are indeed RQ AGN. 
The intrinsic $L_{2-10 ~ keV}$ X-ray luminosities derived using the $L_{1.4 ~ GHz}-L_{2-10 ~ keV}$ relation were then converted into $L_{0.5-2 ~ keV}$ luminosities assuming an intrinsic X-ray spectrum with slope $\Gamma=1.9$. Instead, to derive the intrinsic $L_{0.5-2 ~ keV}$ luminosities for the population of RL AGN selected in Sect.~\ref{sec:RLAGN}, we used the fundamental plane relation derived in \cite{Wang24_FP}. This relation was derived considering a wide sample of RL AGN selected among the deepest known radio fields (GOODS-N, GOODS-S, COSMOS) and extending between $1<z<4$. Following \cite{Wang24_FP} we computed the black hole masses of the RL AGN using the black hole mass to total stellar mass relation reported in \cite{Greene20}, and then we computed the expected intrinsic X-ray luminosity. To finally derive an estimate of the minimum column density obscuring the X-ray emission of the radio-excess AGN candidates (both RQ and RL), we used the mock catalogs presented in \cite{marchesi20}. This catalog contains 5.4M sources simulated down to very faint X-ray fluxes, $\rm 10^{-20} erg\ cm^{-2}\ s^{-1}$ in the 0.5-2 keV band, over an area of 100 $\rm deg^2$ and with different obscuring hydrogen column densities ($20.5<\log (N_H/\rm cm^{-2})<25.5$).  The properties of the sources have been extracted by resampling the X-ray luminosity function of unabsorbed AGN given by \cite{hasinger05}, scaled up by a luminosity–dependent factor to account for the whole AGN population \citep[see][]{gilli07}, and including at $z > 2.7$, a decline in the AGN space density as parameterized in \cite{schmidt95}.
The catalog provides the 0.5-2 keV flux and the corresponding intrinsic 0.5-2 keV X-ray luminosity for each source at a given redshift and obscuration level. Therefore, for each source characterized by a given value of $z$, $L_{0.5-2 ~ keV}$, $f_{0.5-2 ~ keV}$ we derived the correspondent lower limit to the value of $\log N_H$ interpolating the data of the mock catalog for the provided values of redshift, intrinsic X-ray luminosity, and observed flux limit.\\
The histogram in Fig.~\ref{fig:NH_hist} shows the distribution of the inferred lower limits to $\log N_H$ for the 145 X-ray non detected radio-excess sources, considering both the distribution of RQ and RL AGN. The median value of the distribution corresponds to $\log (N_H/\rm cm^{-2})\sim 23.7$, with 103 sources over 145 having $\log (N_H/\rm cm^{-2})>23$ and 44 having $\log (N_H/\rm cm^{-2})>24$, thus being CTK AGN candidates. In particular, 34 out of 89 sources at $z>1.5$, and 9 out of 16 sources at $z>2.5$ are CTK candidates. We recall that these values of $\log N_H$ have to be considered as lower limits to the actual obscuration (given that the observed flux is just an upper limit). We note that there is a significant fraction of sources with $\log (N_H/\rm cm^{-2})\sim 25$. For most of them, the interpolation method allowed only solutions of the obscuration larger than $\log (N_H/\rm cm^{-2})= 25.5$. As it is possible to see, RL sources have a distribution of $\log N_H$ much more skewed towards lower values of the obscuring column density compared to the RQ AGN population, with most of them having a lower limit to $\log N_H$ compatible with being unobscured ($\log (N_H/\rm cm^{-2})\sim20$). Indeed, for a fixed radio luminosity, the expected X-ray luminosity of RL AGN is lower than the one estimated for RQ AGN by, on average, $\sim 3$dex \citep{panessa15,Shang11}. In turn a lower intrinsic $L_{0.5-2keV}$ leads to a lower expected value of $\log N_H$.  In Fig.~\ref{fig:NH_hist} we also show the distribution in $\log N_H$ of the X-ray detected AGN sample of the J1030 field, whose obscuring column densities were estimated using the X-ray spectral analysis in \cite{Signorini23}. The obscuring column densities of these sources are distributed with a median value of $\log (N_H/\rm cm^{-2}) \sim 22.5$ that is $\sim 1$ dex lower compared to the median value of the X-ray non detected radio-excess AGN candidates.  

The Table containing the physical properties and the classification of all the radio sources analyzed on the J1030 field is available at this website: \url{http://j1030-field.oas.inaf.it/jvla_1030} 

\begin{figure}[h!]
    \centering
    \includegraphics[width=1\linewidth]{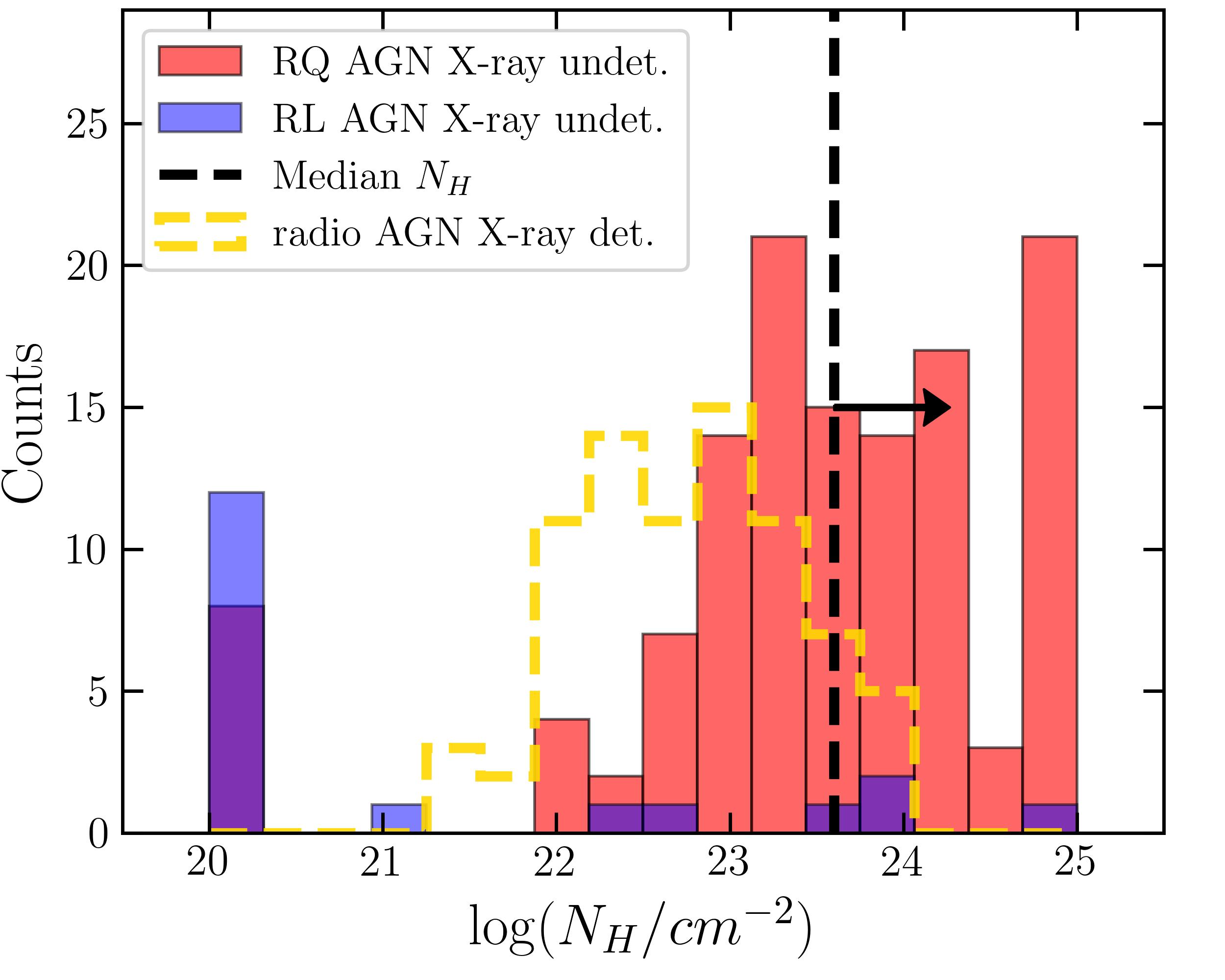}
    \caption{Distribution of the lower limit to the value of $\log (N_H/\rm cm^{-2})$ for the population of radio-excess and X-ray non detected AGN candidates. The red histogram shows the distribution of sources classified as RQ, the blue histogram as RL, while the gold histogram shows the distribution of $\log (N_H/\rm cm^{-2})$ of the X-ray detected radio AGN of J1030 field as derived from the X-ray spectral analysis. The black dashed line marks the median value of the distribution of radio-excess AGN, corresponding to $\log (N_H/\rm cm^{-2})=23.7$.}
    \label{fig:NH_hist}
\end{figure}

\section{Analysis and Discussion} \label{sec:discussion}
 In this Section we discuss and test the reliability of the radio-excess selection performed in Sect.~\ref{sec:Results}, testing the presence of possible contaminants and demonstrating the heavily obscured AGN nature of the radio-excess selected sources. Finally we discuss these results in the context of the cosmic evolution of the population of CTK AGN selected via different methods. 

\subsection{Contamination of the radio-excess sample} 
\label{sec:contamination}
In Sect.~\ref{sec:radioexcessAGN}, we identified a threshold in $REX$ above which the radio emission in excess of what is expected from SF can be reasonably explained assuming an AGN-related additional radio power. However, given the definition of the $REX$ parameter in Eq.~\ref{eq:REX}, there might be contaminants among the radio-excess sources that are not AGN or for which the obscuring column density estimated in Sect.~\ref{sec:obsc} might be overestimated. A detailed analysis of the possible sources of contamination, and of their impact on our selection is provided in Appendix~\ref{app:contamination}, while here we give a summary of the sanity checks in support of the reliability of our selection technique.\\
The total star formation rate ($SFR^{corr}_{SED}$) derived using \texttt{CIGALE} SED fitting is computed by the code by correcting the SF UV-optical emission for dust attenuation since MIR/FIR photometry is not available. This method may underestimate the SFR, in highly dust-obscured galaxies, potentially leading to an overestimation of the $REX$ parameter for some sources, in particular dusty and starburst galaxies. To assess possible contamination from dusty starbursts, we used the 1.1 mm AzTEC image of the J1030 field, where strong dust-obscured starbursts should be detected due to their abundant FIR emission. In addition, we performed a new SED fitting run with \texttt{CIGALE} including a burst component in the fit to check the presence of fainter starburst sources. Both tests suggest that contamination from dust-obscured starbursts among our radio-excess sample is minimal (of the order of few sources).\\
We also investigate whether different SFR timescales can impact the REX classification. While radio-derived SFRs can trace SF histories longer than 150–300 Myr \citep{ArangoToro23}, optical SED-derived SFRs may reflect more instantaneous values. By recalculating REX using average $SFR^{corr}_{SED}$ over the past 100, 300, and 500 Myr, we find no significant shift in the REX distribution, reinforcing the robustness of our selection.\\
Finally, we assess contamination from low-excitation radio galaxies (LERGs), which could be misclassified as obscured AGN due to their radiative inefficient accretion causing a weak radiative output in the optical, UV, and X-ray bands. We used the results from the SED fitting with the AGN component on the radio-excess sample to identify radiatively efficient and radiatively inefficient AGN, following the procedure outlined in \cite{Best23, kondapally22}. We identified 52 LERG candidates, comprising $\sim40\%$ of radio-excess sources at $z<1$, but dropping to $\sim14\%$ at $z>1.5$ and $\sim10\%$ at $z>2.5$. However, the LERGs identification procedure (whose details are described in Appendix~\ref{app:contamination}) led to the inclusion of 7 X-ray detected sources with $\log (L_{2-10\text{ keV}}/\text{erg s$^{-1}$}) > 43$ erg s$^{-1}$ (that are actually radiatively efficient), suggesting that the LERGs classification might be overestimated (given that these sources represent the $\sim 15\%$ of the total LERGs sample). The X-ray stacking analysis described in the next section further confirms that LERGs contamination in our AGN sample is likely largely limited.

\subsection{X-ray stacking analysis} \label{sec:Xstack}

\begin{figure}[h!]
    \centering
   \includegraphics[width=\linewidth]{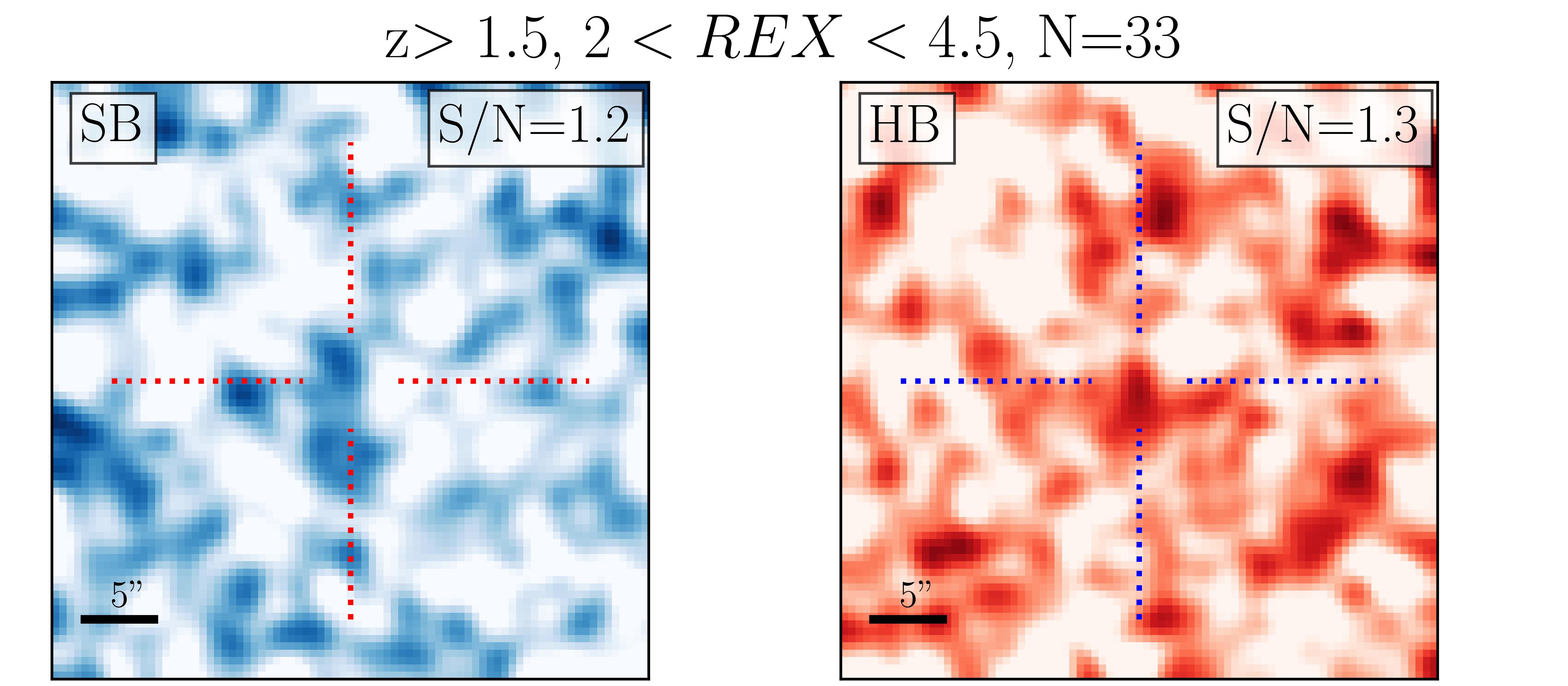}\\
   \includegraphics[width=\linewidth]{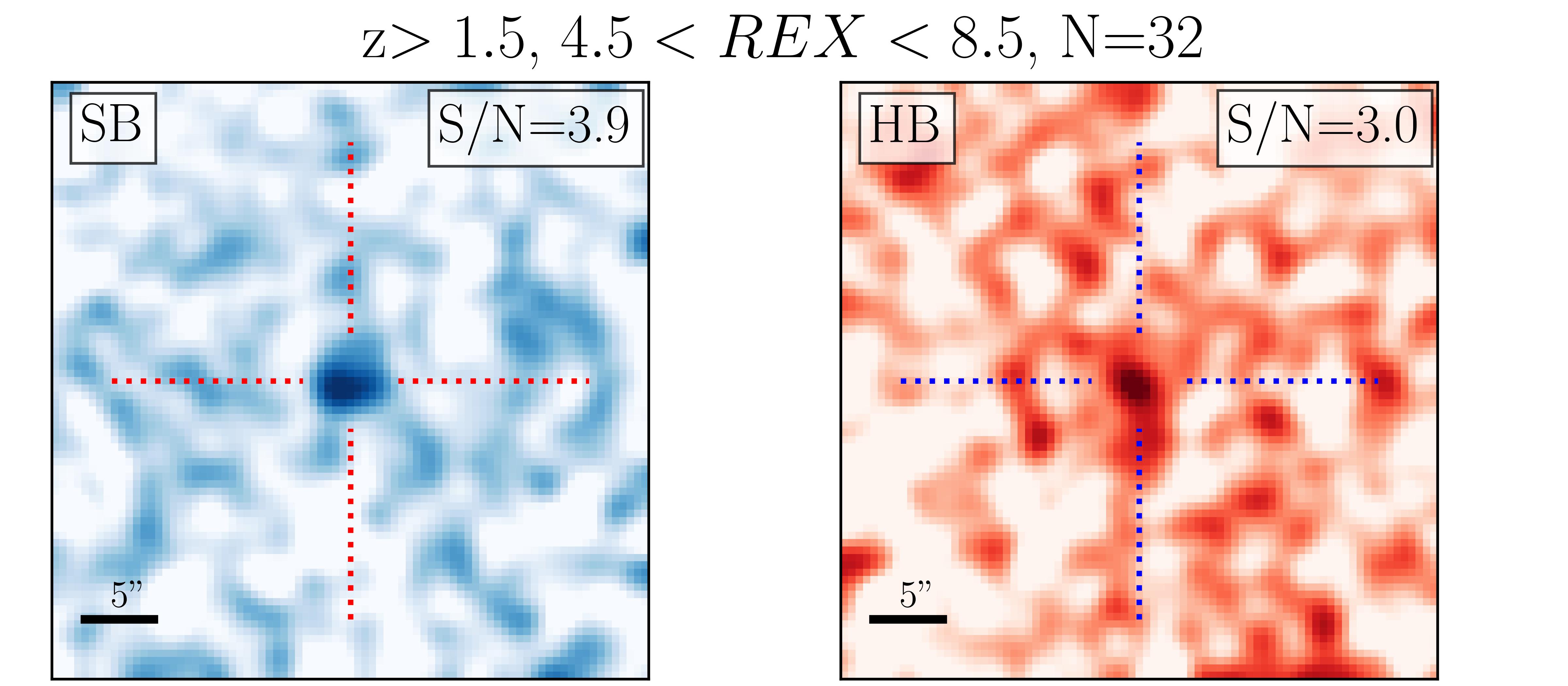}\\
   \includegraphics[width=\linewidth]{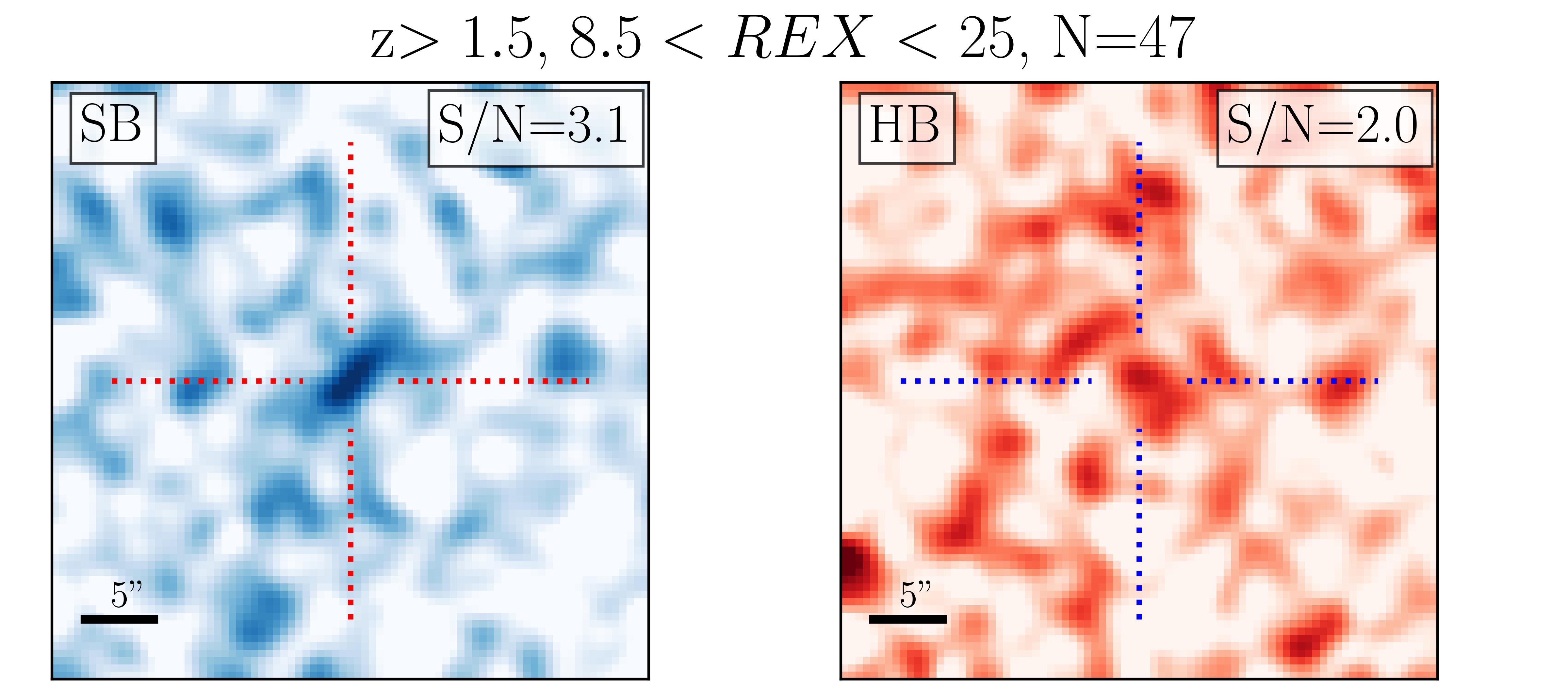}\\
   \includegraphics[width=\linewidth]{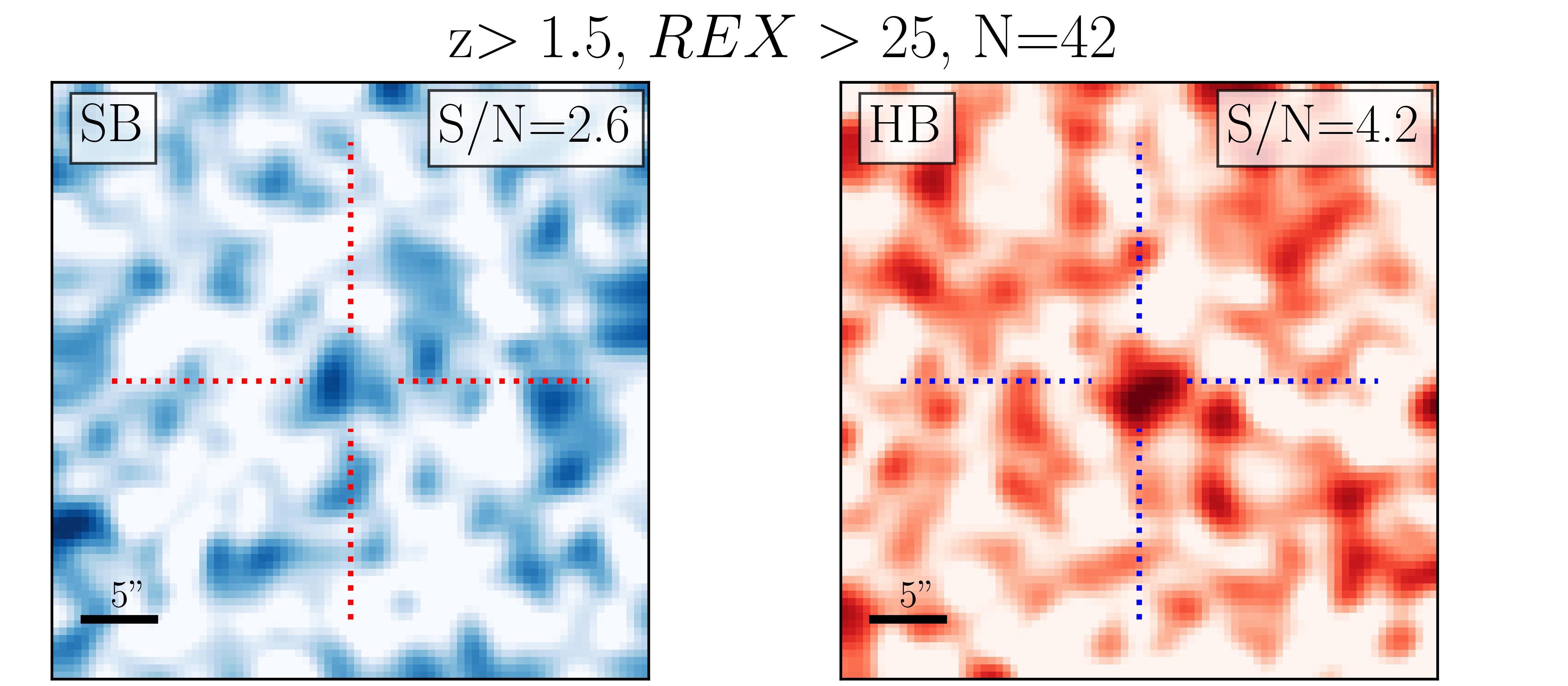}\\
   \includegraphics[width=\linewidth]{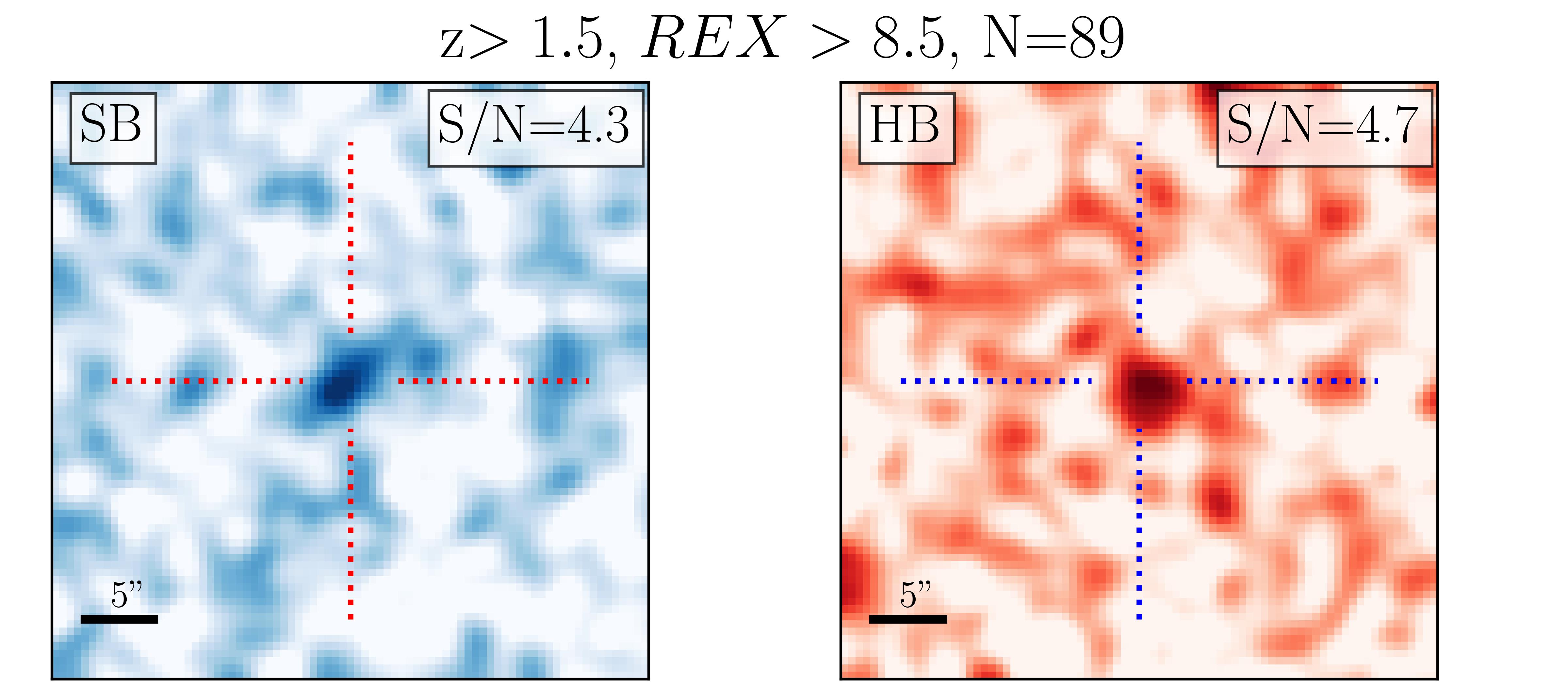}
      \caption{40"$\times$40" X-ray stacking cutouts in the 0.5-2 keV (left) and 2-7 keV (right) of different samples of sources based on their $REX$, from the top to the bottom: $2<REX<4$, $4<REX<8.5$, $8.5<REX<25$, $REX>25$, and $REX>8.5$. The stack was performed considering only sources at $z>1.5$, and excluding those at an off-axis angle $\Theta \lesssim 8^{\prime}$. The stack images are presented using a Gaussian smoothing function. In the upper right corner we show the S/N of the detection  \citep[computed using ][]{Lin83}.}
         \label{fig:Xstack}
\end{figure}
To further prove the goodness of our selection in terms of heavily obscured AGN, we performed an X-ray stacking analysis on the radio excess selected sources falling in the \textit{Chandra} footprint and without an X-ray counterpart.\\
To perform the X-ray stack, we masked all the X-ray detected sources and also sub-threshold detected sources (for a total of 344 sources), and we filled their locations ($6^{\prime \prime}$ for all the sources) with a Poissonian resampling of the local background. Then, we used the \textit{Chandra} tool \texttt{CIAO}\footnote{version 4.15, https://cxc.cfa.harvard.edu/ciao/} to derive the X-ray images corresponding to the mean, and median stack. To derive the count rates and the S/N of each stack we used the following approach. For each of the sources involved in the stack, we define a circular source extraction region with the size determined by the 90\% encircled counts fraction (ECF) radius ($r_{90}$) (fixing a minimum of $1^{\prime \prime}$ and a maximum of $8^{\prime \prime}$). We set the background region for each source to be an annulus with external radius $20$ arcsec and internal radius $r_{90}+1^{\prime \prime}$. The background was computed from the cleaned X-ray image described above. Sources at separation lower than $d<6^{\prime \prime}+r_{90}$ from X-ray detected sources in the original image were excluded from the stack to reduce the noise in the stacking procedure.\\

With the X-ray stacking we are interested in revealing the buried X-ray emission of the population of heavily obscured AGN, but, as reported in Fig.~\ref{fig:NH_hist} there are $\sim 30$ sources for which the level of obscuration estimated from the X-ray upper limit is low ($\log (N_H/\rm cm^{-2})< 22$).  Almost all of these sources are distributed in the outer parts of the X-ray image (i.e. at an off-axis angle $\Theta \gtrsim 8^{\prime}$), where the X-ray flux limit is shallower, therefore we excluded them from the stack. 
Furthermore, we considered only sources at $z>1.5$ because we showed in Appendix~\ref{app:contamination} that the fraction of LERGs possibly contaminating the radio excess AGN population is low ($<14\%$), and also because we want to limit the analysis to the redshift range where the population of heavily obscured AGN is more unconstrained (i.e. at higher redshifts).\\

 We estimated the stacked source count rate using a bootstrap resampling technique. For each source, we considered the individual source count rate and the corresponding effective exposure time. We generated 500 bootstrap realizations by randomly resampling, with replacement, the original set of sources. For each realization, we computed the exposure-time–weighted mean count rate, where the weights are given by the individual exposure times, in order to properly account for variations in sensitivity across the sample \citep[see also][]{miyaji08}. The distribution of the bootstrap realizations was then used to estimate the stacked count rate and its uncertainty. Specifically, we adopted the median of the bootstrap distribution as the best estimate of the stacked count rate, while the uncertainty was derived from the difference between the 50th and 16th percentiles, corresponding to a $1\sigma$ confidence interval.
Following this approach, we derived the count rates in the soft band (0.5-2 keV, SB) and in the hard band (2-7 keV, HB) in four similarly populated $z>1.5$ $REX$ bins: $2<REX<4.5$ (33 sources), $4.5<REX<8.5$ (32 sources), $8.5<REX<25$ (47 sources), $REX>25$ (42 sources).  Then, we derived the S/N of the detection in each band using Eq.~17 in \cite{Lin83}. \\
In Fig.~\ref{fig:Xstack}, we show the X-ray stacking cutouts in the 0.5-2 keV and 2-7 keV for the four different samples. 
The non-detections in the stacks of sources with $2<REX<4.5$ can be possibly explained considering that these sources are mainly normal SFG (lying on the main sequence, see Fig.~\ref{fig:SFR_MS}), and the statistic does not allow the detection of the X-emission coming from SF. We can estimate the X-ray luminosity due to the SFR of these sources using Eq.~15 in \cite{Lehmer16}:
\begin{equation}\label{eq:SFRX}
    L_{2-10 ~ keV,gal}= a_0(1+z)^{\gamma_0}M_{\star} + b_0(1+z)^{\delta_{0}}SFR\ \ \rm erg\ s^{-1}, 
\end{equation}
where (log$a_0$, log$b_0$, $\gamma_0$, $\delta_0$) = (29.30, 39.40, 2.19, 1.02). Considering the median redshift, stellar mass and $SFR^{corr}_{SED}$ of the $2<REX<4.5$ sample we derived an X-ray luminosity of $\sim 5\times 10^{41} \rm erg s^{-1}$, lower than the threshold derived from the $3\sigma$ count rate upper limit ($\sim 10^{42}\rm erg\ s^{-1}$).  On the contrary, sources with $4.5<REX<8.5$, that are clearly detected in both the 0.5-2 keV and 2-7 keV, are probably galaxies that host dust-obscured SF, producing the soft X-ray emission, with also a non-negligible AGN contamination (given that the $3\sigma$ radio-excess selection has to be considered conservative in terms of AGN completeness). From the 0.5-2 keV count rates, we derived an X-ray luminosity of $ 10^{42}\rm erg\ s^{-1}$, that is twice the X-ray luminosity estimated from the $SFR^{corr}_{SED}$ of these sources, but half the one estimated from $SFR_{1.4~GHz}$. This suggests that $SFR^{corr}_{SED}$ might be underestimated for a subsample of sources, but also that part of the radio emission cannot be attributed to SF, but instead to AGN activity. Instead, the lack of detection in the X-ray stacking of the sources with $8.5<REX<25$ is probably due to the combined effect of AGN obscuration and limited statistics, as we will see later. The $S/N=4.2$ detection of the sample of radio-excess sources with $REX>25$ strongly suggests that this sample is dominated by heavily obscured, possibly CTK, AGN, as we will demonstrate in the next Section.\\
To check that the radio-excess selection threshold $REX\geq 8.5$ does not fail in selecting heavily obscured AGN when X-ray non detected sources are considered, we performed a final X-ray stacking analysis merging the last two samples, therefore considering all sources with $REX>8.5$ (but keeping the constraint $z>1.5$).
The X-ray stacks of these 89 sources are shown in the last two panels of Fig.~\ref{fig:Xstack} and confirm a clear detection in the 2-7 keV, strongly supporting the presence of a heavily obscured AGN-dominated population among these sources.

\subsection{X-ray obscuration}\label{sec:Xrayobsc}
\begin{figure}
    \centering
    \includegraphics[width=1\linewidth]{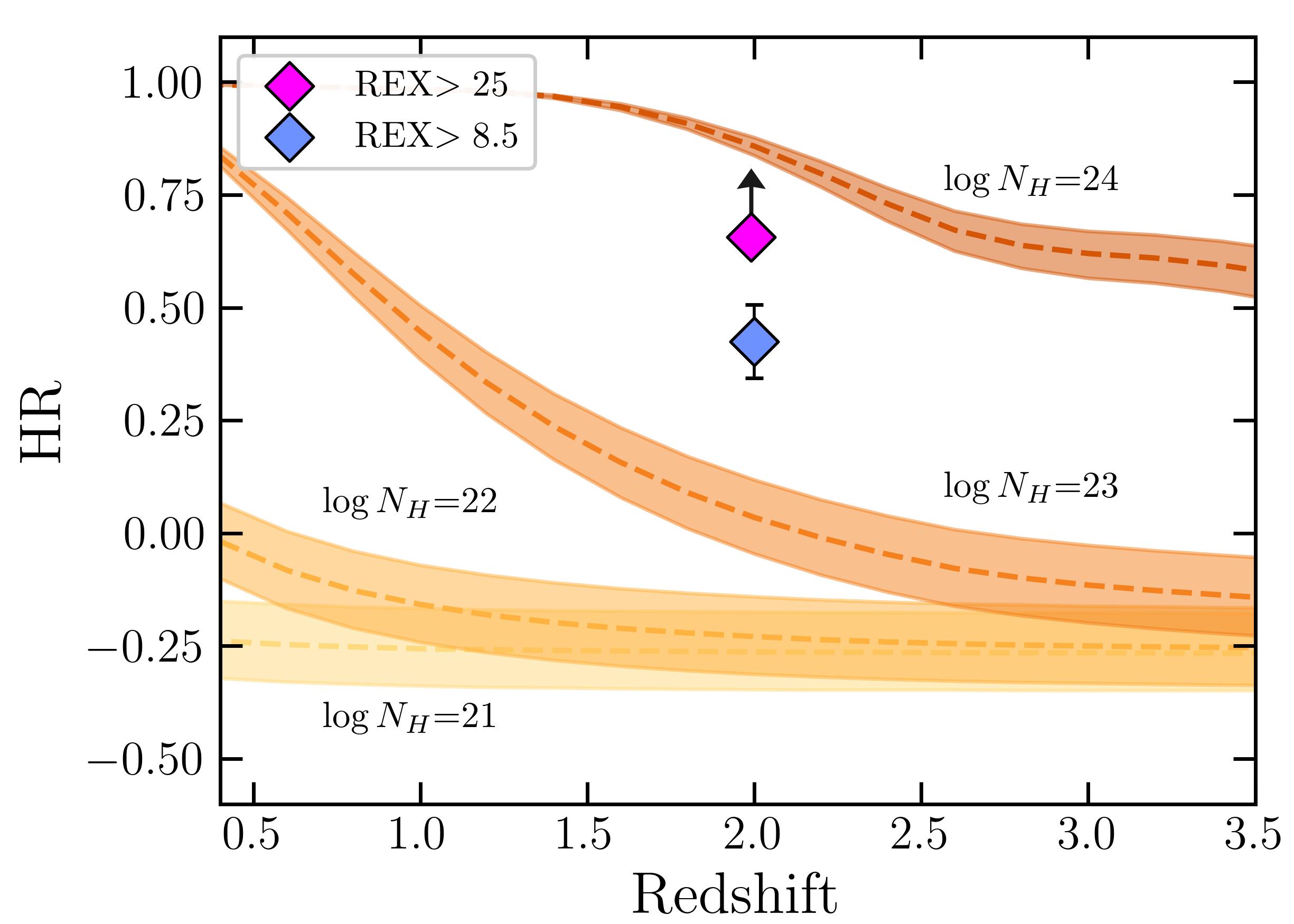}
    \caption{HR versus redshift distribution of the two main samples of radio excess sources at $z>1.5$, identified by $REX>8.5$ and $REX>25$. The shaded areas indicate HR values derived with fixed $\Gamma=1.7$ (top curves) and $\Gamma = 2.1$ (bottom curves), while the dotted lines represent models with $\Gamma= 1.9$, our baseline model. We used the response matrices at the aimpoint of the \textit{Chandra} observations at Cycle 17 \citep{peca21}.}
    \label{fig:HR}
\end{figure}
The X-ray stacking analysis presented in Sect.~\ref{sec:Xstack} suggests that the radio excess selection is really effective in uncovering obscured AGN, especially when it investigates sources non detected in deep X-ray observations, such as those of the J1030 field.  

To further support the heavily obscured AGN nature of these radio-excess selected sources we computed the hardness ratio (HR) of the stacked samples described in Sect.~\ref{sec:Xstack}. This parameter provides an indication, given the median redshift of the sample, of its average obscuration level, and is defined as:
\begin{equation} \label{eq:HR}
    HR=\frac{H-S}{H+S},
\end{equation}
where $S$ and $H$ are the net counts (i.e. background subtracted) in the 0.5-2 keV and 2-7 keV, respectively.\\
In Fig.~\ref{fig:HR} we show the distribution of the HR derived in Sect.~\ref{sec:Xstack} for the samples of $REX>8.5$ and $REX>25$ and presented as a function of the median redshift of each sample (that is $z\sim 2$ for both). In Fig.~\ref{fig:HR} we also show HR for typical AGN column densities ($21<\log (N_H/\rm cm^{-2})<24$) as a function of redshift and with a canonical photon index $\Gamma=1.9\pm 0.2$ \citep{peca21}. The sample of sources with $REX>25$ and with $REX>8.5$ shows HRs that correspond to $23<\log (N_H/\rm cm^{-2})<24$, with the first approaching $\log (N_H/\rm cm^{-2})\sim 24$.\\
We further checked the average obscuration of the two samples using \texttt{Xspec} with the count rates derived in the 0.5-2 keV and 2-7 keV from the stacks. Given the central symmetry of the \textit{Chandra} observations of the J1030 field, we extracted the Ancillary Response File (ARF), and Response Matrix File (RMF) at the average off axis angle of the sources involved in the stack and for each OBSID using the \texttt{specextract} tool and combined them using \texttt{combine\_spectra} tool in the \texttt{CIAO} package\footnote{\url{https://cxc.cfa.harvard.edu/ciao/}}. Then, we used \texttt{Xspec} to derive the obscuration. In particular, we employed a \texttt{MyTORUS} model, which self-consistently accounts for photoelectric absorption, Compton scattering, and fluorescent line emission from a toroidal reprocessor of neutral gas surrounding the central emission \citep{Murphy09,Yaqoob12}. The model includes the line-of-sight absorbed power-law continuum, the Compton-scattered continuum, and the associated Fe K$\alpha$ line emission, all linked through a common photon index ($\Gamma=1.9$) and normalization. The fraction of the transmitted component was fixed to 2.5\%, consistent with most observational results (refs). We used a forward modeling approach with the \texttt{fakeit} command in \texttt{Xspec}, using the ARF and RMF extracted above, to obtain the model count rates in the 0.5-2 keV and 2-7 keV bands by varying the column density of the model and the normalization, while freezing all the other parameters. With this procedure we obtained $\log (N_H/\rm cm^{-2})>23.47$ for the $REX>8.5$ sample and by $\log (N_H/\rm cm^{-2})>23.95$ for the sample at $REX>25$ (where the upper limits come from the non detection in the 0.5-2 keV). Using the normalization obtained from \texttt{MYTORUS}, we then considered the intrinsic (unabsorbed) power-law component alone to compute the intrinsic X-ray fluxes and the corresponding rest-frame 2–10 keV luminosities. We found 2-10keV intrinsic fluxes of $2.16\times 10^{-16}$erg s$^{-1}$ cm$^{-2}$ and $8.65\times10^{-16}$erg s$^{-1}$ cm$^{-2}$ for the $REX>8.5$ and $REX>25$ sources, that correspond to X-ray luminosities of $L_{2-10keV}=5.9\times 10^{42}$erg s$^{-1}$ and $L_{2-10keV}=2.3\times 10^{43}$erg s$^{-1}$, respectively.

These X-ray luminosities are too high to be justified only by SF. The average SFR returned by \texttt{CIGALE} for sources at $REX>8.5$ and $REX>25$ are 18 $M_{\odot}yr^{-1}$ and 5 $M_{\odot}yr^{-1}$, respectively, while those returned by the radio luminosity are both $\sim 300 M_{\odot}yr^{-1}$. Considering the average stellar masses of the two samples we derived the correspondent X-ray luminosity using Eq.~\ref{eq:SFRX}. For both samples, the predicted X-ray luminosities considering $SFR^{corr}_{SED}$ are $<2\times 10^{41}$erg s$^{-1}$, while taking $SFR_{1.4 GHz}$ the expected X-ray luminosities are $<2\times 10^{42}$erg s$^{-1}$. In both cases, they are lower than the average intrinsic X-ray luminosities derived from the X-ray analysis.\\
Finally, we note that the X-ray luminosities obtained from the stack of $REX>8.5$ and $REX>25$ sources strongly support a very limited contamination from LERGs among the radio-excess AGN because LERGs are not expected to be characterized by these level of X-ray emission.\\
Another class of sources that can contribute to the amount of radio detected obscured AGN are the 99 radio sources that are not detected in the Ks-band selected catalog of the J1030 field (see Sect.~\ref{sec:J1030rad}). Using X-ray stacking we indeed show in Appendix~\ref{app:Xstack_radiound} that at least part of this sample can be made by heavily obscured AGN.

\subsection{Comparison with radio predictions} \label{sec:model}
\cite{Mazzolari24} developed an analytical model to predict the number of radio-detectable AGN over a given radio image considering populations of AGN at different levels of obscuration. In particular, they extrapolated the predictions of the AGN population synthesis model of the CXB to the radio band by deriving the 1.4 GHz luminosity functions of unobscured, obscured, and CTK AGN. Then, they used these functions to forecast the number of detectable AGN based on the area, flux limit, and completeness of a given radio survey and compared these results with the AGN number resulting from X-ray predictions. They found that, while X-ray selection is generally more effective in detecting unobscured AGN, the surface density of CTK AGN returned by the radio selection is $\sim 10$ times larger than the X-ray one, and even greater at $z>3$. Since the models in both bands assumed the same intrinsic population of AGN, the difference in detectability of heavily obscured AGN was a direct consequence of the fact that obscuration has no effect on radio emission. In that work, the authors also considered the J1030 field, predicting over the whole radio image ($\sim0.2$deg$^2$) 533 radio detectable AGN, of which 222 CTK. Since in our analysis, we are only considering sources in the X-ray image footprint ($\sim 0.09$deg$^2$), the expectations for the intrinsic radio AGN population of \cite{Mazzolari24} have to be rescaled. However, as mentioned above, the model in \cite{Mazzolari24} was based on the intrinsic population derived from the CXB model and, therefore, from X-ray observations, but it is possible that the intrinsic AGN population might be larger, in particular at $z>3$, as suggested by recent JWST results \citep{yang23, Akins24}.\\
In the $0.09$deg$^2$ of the J1030 field covered by the X-ray image our radio-excess selection returned 181 radio-excess sources, of which 145 are not X-ray detected and 44 have an estimated $\log (N_H/\rm cm^{-2})>24$. These numbers are lower than the 239 and 100 radio AGN and CTK AGN predicted by the model in the same area. However, we have to recall that our selection is based only on radio-excess criteria, and different works showed that the radio-excess selection techniques can select only a fraction of the total radio AGN population, ranging between $30-60\%$ depending on the depth of the radio observations, on the availability of multiband information and on the threshold chosen for the radio excess selection \citep{Mazzolari24, zhu23, smolcic17b, bonzini13}. Therefore, before computing the number density of radio-excess selected CTK AGN, we have to estimate the statistical completeness corrections related to our radio excess selection. 

\subsection{Radio excess completeness corrections}\label{sec:corrections}
Based on our data, we can follow two different approaches to estimate the incompleteness of our radio-excess selection. The first exploits the distribution of radio sources as a function of the $REX$ parameter. As it is possible to see from Fig.~\ref{fig:REXhist}, there is an excess of radio sources with respect to the Gaussian distribution describing the SFG population below the radio-excess threshold. This excess can be ascribed to the presence of AGN with a nuclear radio contribution not strong enough to push them above the $3\sigma$ radio excess threshold. This scenario is further supported by the results of the X-ray stacking discussed in Sect.~\ref{sec:Xstack}, where we showed that also at $REX<8.5$, there might be a nonnegligible contribution from AGN. In principle, these sources might also be SFG with larger IR emission (and therefore larger radio emission given the FIRC), but this population should already be included in the Gaussian distribution (given its dispersion around the peak). The number of sources in excess to the SFG Gaussian distribution and that are between the peak of the distribution and the $REX$ threshold corresponds to $\sim70\%$ of the sources identified as radio-excess. This leads to a correction factor of 1.7 (i.e. 0.25 dex) to the number of AGN derived considering only the $3\sigma$ radio-excess selected sources. 
Another way to estimate the statistical incompleteness is by looking at the population of X-ray detected radio AGN. As it is possible to see from the cumulative distribution of these sources in Fig.~\ref{fig:REXhist}, even if the X-ray AGN distribution is significantly skewed towards larger $REX$ values, only $\sim$40\% of the X-ray detected AGN are selected as radio-excess. This means that X-ray detected radio AGN below the radio-excess threshold are 1.5 times more abundant than radio-excess ones, leading to a correction factor to the observed quantities of $\sim 2.5$ (i.e. 0.4 dex).\\
While the second approach allows us to correct also for sources whose nuclear radio emission is completely overwhelmed by the host galaxy SF (i.e., sources having REX$<1$), the first approach provides a completeness correction based only on the radio-excess parameter and is, therefore, more appropriate and conservative.

\begin{figure*}[h]
    \centering
   \includegraphics[width=0.9\linewidth]{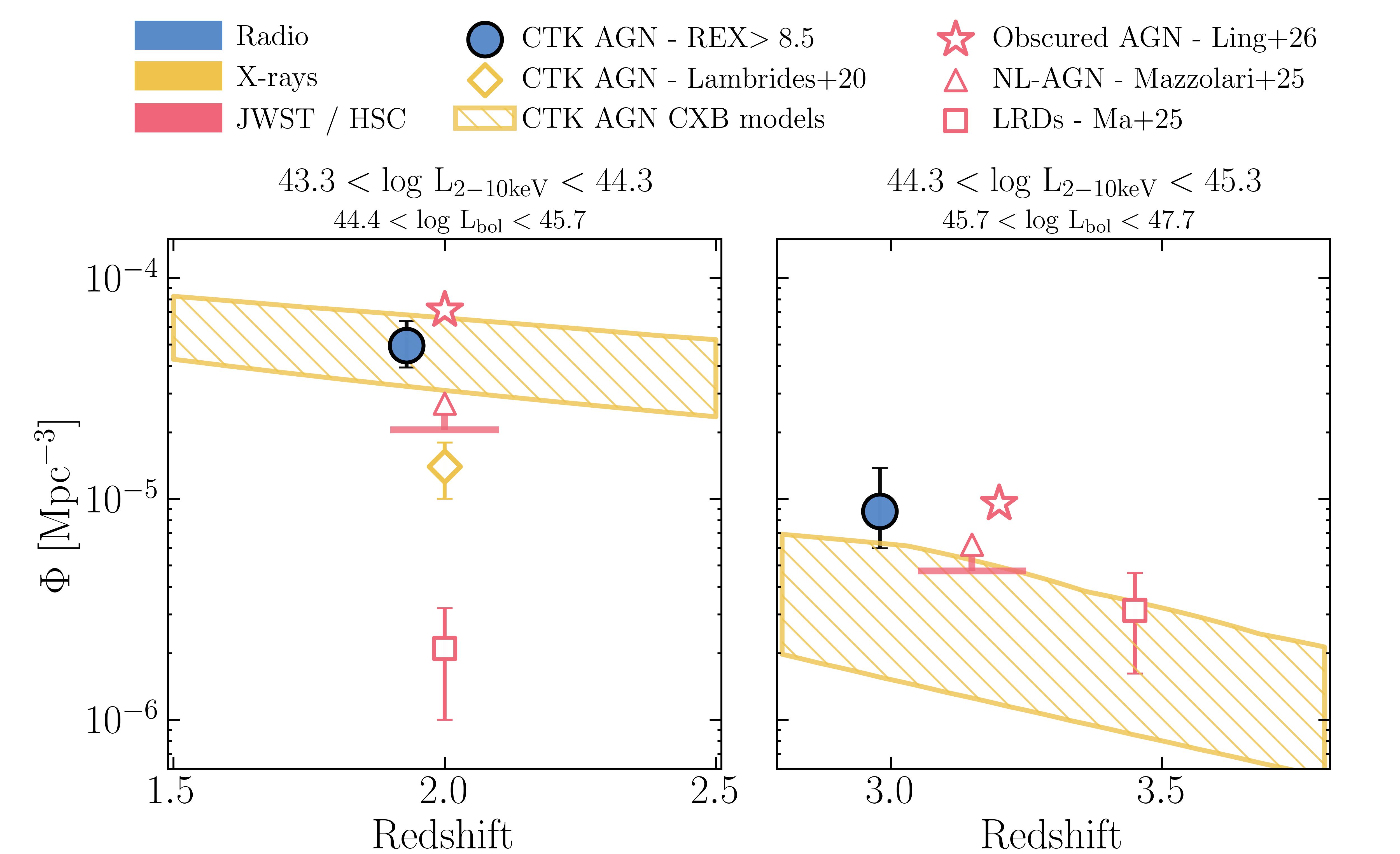}
      \caption{ Comparison between the number densities of radio-selected CTK AGN and the results from X-ray models and observations. The left panel refer to the $z\sim2$ and $43.3<\log (L_{2-10\text{ keV}}/\text{erg s$^{-1}$})<44.3$, the right panel to $z\sim3$ and $44.3<\log (L_{2-10\text{ keV}}/\text{erg s$^{-1}$})<45.3$. We also report the corresponding bolometric luminosity ranges assuming standard AGN bolometric corrections from \cite{duras20}. Blue refers to the radio selected CTK AGN number densities derived in this work (see Sect.~\ref{sec:numberdensity}), yellow to X-ray results, red to recent results considering JWST photometric or spectroscopic data. The upper and lower boundaries of the hatched regions are defined by the CTK AGN number density predicted by the CXB synthesis models of \cite{ananna19} and \cite{gilli07}, respectively. The yellow diamond represents the observed X-ray CTK AGN number density computed in \cite{lambrides20} in the CDFS at $z\sim2$. The red stars mark the obscured AGN number density obtained in \cite{Ling26} from JWST/MIRI observation of the SMILES program, while the red triangles show the lower limit to the obscured AGN number density derived in \cite{Mazzolari25} selecting narrow-line AGN among the spectra of the JWST CEERS survey. Finally, the squares represent the LRD number density derived in \cite{Ma25}. Data points are plotted in correspondence of the median redshifts of the samples. }
         \label{fig:ND}
\end{figure*}

\subsection{Compton-thick AGN number density and fraction} 
\label{sec:numberdensity}
\subsubsection{Radio excess number density}
To compute the AGN number density we followed the same approach reported in \cite{vito14}, starting from the binned luminosity function \citep{Page00}. The binned luminosity function for a given redshift and luminosity bins containing N sources is given by:
\begin{equation}\label{eq:binnedLF}
    \phi = \frac{N}{\int_{zmin}^{zmax}\int_{\log Lmin}^{\log Lmax} \Omega\  \frac{dV}{dz}\ dz\ dlog L },
\end{equation}
where $\Omega=\Omega(S)=\Omega(z,L)$ is the fractional sky coverage, that is the fraction of the sky covered by a survey at a given observed flux $S$ corresponding to a source luminosity $\log L$ at redshift $z$, while $dV/dz$ is the comoving volume element :
\begin{align}
   \frac{dV}{dz}=\frac{4\pi c}{H_{\rm 0}} \frac{D^{2}_{\rm M}}{\bigl(\Omega_{\rm m}(1+z)^3 + \Omega_k (1+z)^2 + \Omega_{\rm \Lambda}\bigr)^{1/2}},
\end{align}
where $D_{\rm M}$ is the comoving distance. Then, the number density of sources is returned by multiplying $\Phi$ in Eq.~\ref{eq:binnedLF} by $\Delta \log L$. In particular, we computed the heavily obscured AGN number densities considering two different redshift-luminosity bins: $1.5<z<2.5$ at $43.3<\log (L_{2-10\text{ keV}}/\text{erg s$^{-1}$})<44.3$, and $2.8<z<3.8$ at $44.3<\log (L_{2-10\text{ keV}}/\text{erg s$^{-1}$})<45.3$. The intrinsic X-ray luminosities are those computed in Sect.~\ref{sec:obsc}. The two redshift and luminosity intervals are almost contiguous, and the choice to use the X-ray luminosity as a reference allows us to compare our results with other measurements coming from X-ray observations. The analysis presented in Sect.~\ref{sec:contamination} and Sect.~\ref{sec:Xrayobsc} revealed the reliability of our radio-excess threshold (REX=8.5), and since we are interested in the most heavily obscured AGN population, we considered for both intervals only sources with an estimated $\log (N_H/\rm cm^{-2})>24$ (from those computed in Sect.~\ref{sec:obsc}). 
In the lower redshift-luminosity bin we counted 8 sources, while in the other bin we have 4 sources. We then derived the CTK AGN number densities using Eq.~\ref{eq:binnedLF} and computed upper and lower errors at 68.3\% confidence levels using the standard Gehrels approximation for low count statistics \citep{Gehrels86}. Finally, we applied to the observed number densities the radio-excess completeness correction of 1.7 estimated in Sect.~\ref{sec:corrections}. We obtained $n=4.94\times 10^{-5}\ \rm Mpc^{-3}$ for the $z\sim2$ and $43.3<\log (L_{2-10\text{ keV}}/\text{erg s$^{-1}$})<44.3$ bin, while we got $n= 8.78\times 10^{-6}\ \rm Mpc^{-3}$ for the $z\sim 3$ and $44.3<\log (L_{2-10\text{ keV}}/\text{erg s$^{-1}$})<45.3$ bin (see Table~\ref{tab:number_density_ctk}).

\subsubsection{Comparison with X-ray results}
In Fig.~\ref{fig:ND} we show the CTK AGN number densities derived from our selection compared with other results coming from X-ray observations and models.
In particular, \cite{lambrides20}, considering the same redshift and X-ray luminosity range as in our lowest bin, derived the CTK AGN number density considering deep X-ray observations. In that work, by studying the multiwavelength properties of sources classified as low-luminosity AGN on the \textit{Chandra} Deep Field South \citep[CDFS][]{luo17}, they found that many of these sources were heavily obscured AGN disguised as low-luminosity AGN, with column densities an order of magnitude higher than what was previously derived. In particular, the X-ray number density reported in Fig.~\ref{fig:ND} was derived considering those sources with measured $\log (N_H/\rm cm^{-2})>24$. \\
In Fig.~\ref{fig:ND}, we additionally show for both the X-ray luminosity ranges the trend of the intrinsic CTK AGN X-ray number density predicted by two different population synthesis models of the CXB. In particular, the lower limit to the two shaded regions is given by the CXB model of \cite{gilli07} while the upper boundary is provided by the model of \cite{ananna19} that predicts the largest number of CTK AGN among the other CXB models \citep{aird15,ueda14,buchner15}. For both models, we selected only sources with $24<\log (N_H/\rm cm^{-2})<26$.

In the lowest redshift bin ($z\sim2$), the radio-excess CTK AGN number density is compatible with the range of the intrinsic CTK AGN population returned by the CXB models. In this redshift range the BHARD functions derived from simulations \citep{sijacki15,Volonteri16,shankar14}, infrared observations \citep{delvecchio17}, and X-ray observations \citep{vito18,aird15,ananna19} broadly agree, suggesting that also the population of heavily obscured AGN is well constrained. Therefore, the fact that the radio-selected CTK AGN population is not in tension with the result from the X-ray population synthesis models at $z\sim2$ is not surprising but rather supports the goodness and purity of our selection. At the same time, our observed data point is $\sim$5 times larger than the CTK AGN number density estimated by \cite{lambrides20} using single field deep X-ray data. This supports the results presented in \cite{Mazzolari24}, showing that radio observations can be more efficient than single X-ray observations in unveiling the population of heavily obscured AGN.\\
At $z\sim3$, the radio CTK AGN number density derived in this work is a factor of $2-3$ higher than what is predicted by the CXB models for the same redshift and luminosity range. At $z>3$, the tension between the BHARD function derived from X-ray observation and the larger one predicted by simulations and recently observed in JWST data \citep{yang23,Lyu23} suggests that the population of heavily obscured AGN at these redshifts is likely larger than what observed from X-rays. In this context, our result supports this hypothesis, showing that the most obscured AGN population at high-z can actually be larger than what is expected from X-rays. \\
These results support the advantage of using radio surveys to investigate the heavily obscured AGN population, in particular when coupled with deep X-ray observations (as in the case of the J1030 field), allowing an extremely sensitive search for these sources. We also stress that our radio-excess selected CTK AGN number densities, even if corrected for the statistical incompleteness of the radio-excess selection, have still to be considered conservative as discussed in Sect.~\ref{sec:corrections}. Furthermore, the X-ray stacking performed in Appendix~\ref{app:Xstack_radiound} on the sample of radio sources without a Ks-band counterpart (and therefore not included in our analysis) showed that also these sources can host a consistent fraction of heavily obscured AGN, probably making our number densities even more conservative. We further note that even considering a $\sim 15\%$ contamination of the sample due to the possible presence of LERGs (see Sect.\ref{sec:contamination}) our results remain basically unchanged.

\begin{table}[htbp]
\centering
\caption{Radio selected CTK AGN number densities and fractions in the two luminosity and redshift bin}
\label{tab:number_density_ctk}

\begin{tabular}{ccc}

\hline
\vspace{-0.3cm}\\
$\log (L_{2-10\text{ keV}}/\text{erg s$^{-1}$})$ & $43.3-44.3$ & $44.3-45.3$ \\

$z$ & $1.5-2.5$ & $2.8-3.8$ \\
\vspace{-0.3cm}\\

\hline
\\

Number Density [$\rm Mpc^{-3}$]& $4.94^{+1.4}_{-1.0}\times 10^{-5}$ & $8.78^{+5.0}_{-2.8}\times 10^{-6}$ \\
\\

CTK AGN Fraction & $ 0.40^{+0.11}_{-0.08}$ & $0.59^{+0.34}_{-0.19}$ \\
\vspace{-0.2cm}\\

\hline
\end{tabular}

\end{table}

\subsubsection{Compton-thick AGN fraction}\label{sec:CTK_frac}
 The number densities computed in Sect.~\ref{sec:numberdensity}, especially at $z\sim3$, may point to a larger CTK AGN population than what was measured in other works and than what was expected from CXB synthesis models. Therefore, it is worth quantifying the corresponding CTK AGN fractions relative to the whole AGN population.
We took the total AGN bolometric luminosity function determined by \cite{Shen20}, obtained using an AGN compilation that includes observations in the rest-frame IR, B-band, UV, soft, and hard X-ray from past decades and following the same approach as in \cite{Hopkins07}. This bolometric luminosity function accounts for both obscured and unobscured AGN populations, and we considered it as a baseline for the entire AGN population. Since the whole AGN luminosity function computed in \cite{Shen20} is in terms of AGN bolometric luminosity, to correctly compare these results with our number densities we converted our X-ray luminosity bin into bolometric luminosity bin using standard AGN $\rm L_{bol}-L_{2-10\ keV}$ bolometric corrections from \cite[][reported in the labels in Fig.~\ref{fig:ND}]{duras20}. 

At $z\sim2$ our CTK AGN number density translates into a CTK AGN fraction ($f_{\rm CTK}$) of  $ 0.40^{+0.11}_{-0.08}$. The CTK AGN fraction expected by CXB models at the same redshift-luminosity bin, and calculated with respect to the total AGN population predicted by \cite{Shen20}, ranges from $f_{\rm CTK}\sim 0.25$ in \cite{gilli07} to $f_{\rm CTK}\sim 0.52$ in \cite{ananna19}, 
perfectly consistent with our results. 
Instead, at $z\sim 3$, we found $f_{\rm CTK}= 0.59^{+0.34}_{-0.19}$, whereas the CTK AGN fraction predicted by CXB models ranges from $f_{\rm CTK}\sim 0.10$ in \cite{gilli07} to $f_{\rm CTK}\sim 0.41$ in \cite{ananna19}. Therefore, in this redshift-luminosity bin we found a CTK AGN fraction larger than the average value of the two CXB models (as expected from Fig.~\ref{fig:ND}). 

To investigate whether our results predict a significant evolution of the obscured AGN fraction with redshift, we considered only the high luminosity bin $44.3<\log (L_{2-10keV}/ \rm erg\ s^{-1} )<45.3$ and computed $f_{\rm CTK}$ predicted by the CXB models at $z=2$ in that luminosity bin, which can be directly compared with our result at $z=3$. The CTK AGN fraction from the CXB models at $z=2$ is $f_{\rm CTK}\sim 0.10$ in \cite{gilli07} and $f_{\rm CTK}\sim 0.35$ in \cite{ananna19}, lower than the value $f_{\rm CTK}=0.59$ found by our radio-excess selection at $z=3$. Such an increase of the CTK AGN fraction with redshift \citep[$\Delta f_{CTK, z=2-3}\sim 0.24$, when compared to][]{ananna19} is in line with the recent JWST findings of a larger AGN population not detected in the X-rays. However, we caution that the $\Delta f_{CTK, z=2-3}$ value we found should be considered as an upper limit to the actual increase of the CTK AGN fraction in this redshift range, given that we assumed at $z=2$ the CTK fraction of CXB models, which, as we saw, might have missed part of the CTK AGN population.

\subsubsection{Comparison with recent JWST results}\label{sec:JWST_comp}
In Fig.~\ref{fig:ND} we also report the results on the obscured and heavily obscured AGN selection coming from recent works exploiting JWST photometric and spectroscopic data. In particular, \cite{Ling26} used deep JWST/MIRI images from the Systematic Mid-infrared Instrument Legacy Extragalactic Survey (SMILES) program to select obscured AGN in the GOODS-S field based on their SED-fitting decomposition. Out of the 534 selected AGN, only 37 have an X-ray counterpart in the CDFS, suggesting that most of them should be heavily obscured. Instead, \cite{Mazzolari25}, used JWST/NIRSpec medium resolution spectra from the CEERS program to select narrow-line AGN using several emission line diagnostic diagrams. Also in this case, only 5 out of the 52 selected narrow-line AGN have an X-ray counterpart in the 800ks \textit{Chandra} image covering the EGS field, and all of them have column densities $\log (N_H/\rm cm^{-2})>22.5$. In particular, the results presented in \cite{Mazzolari25} represent only a lower limit to the spectroscopically selected heavily obscured AGN number densities, given the lack of completeness corrections. Given that most of the sources considered in these works do not have any X-ray detection, it was not possible to directly use our X-ray luminosity bin to compute the number densities. We therefore converted the X-ray luminosity range into an AGN bolometric luminosity range using the bolometric corrections from \cite{duras20} and then compared with the AGN bolometric luminosities reported in those works.
We note that our CTK AGN number densities are consistent with the lower limits derived from \citep{Mazzolari25} and also consistently lower than the JWST-selected obscured AGN number densities derived in \cite{Ling26}. This is expected given that the SED-fitting based selection performed in \cite{Ling26} probably also includes obscured AGN that are not necessarily CTK. Additionally, we note that the ratio between the number density from \cite{Ling26} and the CTK AGN one derived in this work decreases between $z \sim 2$ and $z\sim 3$, suggesting that the abundance of CTK AGN with redshift might increase, as already outlined in Sect.~\ref{sec:CTK_frac}. 

Finally, we compared our results with the $2<z<3$ LRDs number densities from \cite{Ma25}. LRDs are peculiar Type I AGN, showing broad Balmer lines (associated with broad line region emission), compact morphologies, and a peculiar V-shape SED in the rest-frame UV and optical \citep{Matthee23, Greene23, Akins24, Kocevski23, Kokorev24, Hviding25}. These sources appear to populate in particular the $z>5$ Universe, typically have bolometric luminosities $\log (L_{bol}\rm / erg s^{-1})<46$ \citep[lacking high-luminosity counterparts][]{Ma25_LRDQSO} and given their redshift distribution and physical properties it has been hypothesized that they should represent one of the first episodes of accretion onto SMBHs \citep{Cenci25_LRD_DCBH,Pacucci25_LRDnurseries}. One of their peculiar properties is the lack of X-ray emission, with observed X-ray luminosity upper limits up to two or three orders of magnitude lower than their expected intrinsic luminosities, which have been suggested to arise from CTK gas envelopes around the central SMBHs \citep{Inayoshi25,Maiolino24_X} or from an intrinsic X-ray weakness \citep{Lambrides24, Madau24}. It is also important to note that this population has not yet been detected in any deep radio observations \citep[even with stack, see][]{Akins24,Gloudemans25,Mazzolari26}, but this might be due to sensitivity limitations that could be overcome by future radio facilities such as SKAO \citep{Latif24,Mazzolari2026.SKA}. \cite{Ma25}, using Hyper Suprime Camera (HSC) images, looked at $z\sim 2$ analogs of these LRDs and derived the number density evolution reported in Fig.~\ref{fig:ND} \citep[see also][for a similar estimate]{Loiacono25}. The LRD number density at $z\sim 2$ is 1.3 dex lower than the CTK AGN number density we derived, while at $z\sim 3$ (and in the upper luminosity bin) it is a factor of three lower than our findings. The distinct trend with redshift of the LRD population highlights that these sources should not be (only) standard CTK AGN, and that there must be other properties driving their peculiar evolution.

\section{Conclusions}\label{sec:conclusions}
In this work, we studied the population of radio sources of the J1030 field looking for heavily obscured AGN. We started from the 1003 radio sources with a counterpart in the Ks-band selected multiwavelength catalog, 
and we performed a radio-excess selection to identify AGN candidates. Then we focused on those that are non detected in the deep \textit{Chandra} image to select only extremely obscured sources.\\
The results can be summarized as follows:
\begin{itemize}
    \item We defined a radio excess parameter $REX$ that corresponds to the ratio between the $SFR_{1.4 GHz}$ computed directly from the radio luminosity of the sources and the $SFR^{corr}_{SED}$ derived from the SED fitting to the optical and NIR photometry (and corrected for dust attenuation). Large values of $REX$ indicate that there is radio emission in excess with respect to what is expected from pure SF. Fitting the $REX$ distribution with a Gaussian, we found the ($3\sigma$) threshold to identify radio-excess sources in correspondence of $REX\sim 8.5$. Among the 1003 radio sources with a Ks-band counterpart, we identified 233 radio-excess AGN candidates.
    \item We then focused on the radio-excess sources located in the deep X-ray image footprint but without an X-ray detection (145 sources). We first estimated a lower limit to the obscuring column density using the observed X-ray flux limit and the estimated intrinsic X-ray luminosity, finding a median value of $\log (N_H/\rm cm^{-2}) >23.7$. Then, by performing a detailed X-ray stacking analysis, we found solid indications that this population is dominated by heavily obscured AGN. In particular, the stacks of samples with $REX>8.5$ and $REX>25$ show clear detections in the 2-7 keV (S/N$>4$), and the obscurations, inferred by using the observed HRs, are compatible with mildly CTK values. Furthermore, the median X-ray luminosities of the sources in the two samples are too high to be justified only by SF, supporting the reliability of our selection.
    \item We finally estimated the radio-excess selected CTK AGN number densities in two different redshift-luminosity bins one at $z\sim2$ and the other at $z\sim3$ by selecting only sources with estimated column densities larger than $\log (N_H/\rm cm^{-2})>24$. We stress that this is the first time that the number density of CTK AGN is estimated from a radio perspective. At $z\sim2$ our results are compatible with the expectations for the CTK AGN  derived from CXB models, but are much larger than the X-ray number density derived considering single deep X-ray observations. On the contrary, at $z>3$ our number density is 2-3 times larger than that predicted by the CXB models and more in line with the results from simulation and recent JWST results.  At $z\sim 3$ we found a CTK AGN fraction $f_{\rm CTK}\sim 0.6$, larger than the one predicted by the X-ray models, and marking a clear increase in the CTK AGN fraction between $z\sim2$ and $z>3$.
    \item The radio-excess selection of heavily obscured AGN and the results summarized above are robust, as demonstrated in details in Appendix~\ref{app:contamination} where we investigate the possible sources contaminating our selection. In particular we showed that our radio-excess selected sample is not significantly contaminated neither by dust obscured starburst galaxy, nor by LERGs.
\end{itemize}
This work shows the importance of finding new and complementary ways to select the most obscured AGN, especially at high redshift. In particular, this work shows the advantages of combining the information coming from deep X-ray and radio observations to unveil this elusive population. The campaigns that will be carried out by future radio and X-ray facilities, like the Square Kilometer Array Observatory (\textit{SKAO}), \textit{NewAthena} or the Advance X-ray Imaging Satellite (\textit{AXIS}), will allow expanding this kind of analysis to larger samples and multiple fields, reaching unparallel depth both in radio and X-rays.\\
The confirmation of the heavily obscured AGN nature of the selected sources will be further proved by an extensive spectroscopic follow-up that we are carrying out with LBT/MODS and ESO/VLT (PI: G. Mazzolari) and that will provide rest-frame UV and optical spectra of more than 200 sources in the J1030 field, the large majority of which will be radio-excess selected galaxies. In this way, we will have the possibility not only to confirm their nature but also identify any peculiar spectroscopic feature characterizing these radio-selected AGN.

\begin{acknowledgements}
GM acknowledges useful discussions with Gianni Zamorani and Matteo Sapori. GM acknowledges the support from Centro Universitario Cattolico (CUC). We acknowledge financial support from the INAF Grants for Fundamental Research 2023 and from the grant PRIN MIUR 2017PH3WAT (‘Black hole winds and the baryon life cycle of galaxies’). GM acknowledges funding by the European Union (ERC APEX, 101164796). Views and opinions expressed are, however, those of the authors only and do not necessarily reflect those of the European Union or the European Research Council Executive Agency. Neither the European Union nor the granting authority can be held responsible for them. KI acknowledges support under the grant PID2022-136827NB-C44 provided by MCIN/AEI/10.13039/501100011033 / FEDER, UE. LF acknowledges support from the INAF 2023 mini-grant "Exploiting the powerful capabilities of JWST/NIRSpec to unveil the distant Universe". IP acknowledges support from INAF under the Large Grant 2022 funding scheme (project "MeerKAT and LOFAR Team up: a Unique Radio Window on Galaxy/AGN co-Evolution"). MS acknowledges financial support from the Italian Ministry for University and Research, through the grant PNRR-M4C2-I1.1-PRIN 2022-PE9-SEAWIND: Super-Eddington Accretion: Wind, INflow and Disk-F53D23001250006-NextGenerationEU. MS also acknowledges support through the European Space Agency (ESA) Research Fellowship Programme in Space Science.
\end{acknowledgements}

%
%

\bibliographystyle{aa}
\bibliography{literature}

\begin{appendix}

\section{Photo-z quality assestment for the J1030 multiband catalog}
We report here the values of the photoz quality estimators obtained from the comparison of the photometric and spectroscopic redshifts of the 219 sources used for the test.
\begin{table}[h!]
    \centering
    \begin{tabular}{c|ccc}
        Code-Template & $\sigma_{NMAD}$& $\eta$ [\%] & $\eta_{rel}$ [\%]   \vspace{0.15cm}\\
        \texttt{eazy}-'tweak' & 0.044 & 16.9 & 21.5  \\ 
         \texttt{eazy}-'sfhz' & 0.051 & 18.3 & 18.3  \\
         \texttt{LePhare}-galaxy & 0.085 & 23.7 & 13.7  \\
         \texttt{LePhare}-AGN & 0.088 & 41.1 & 20.5  \\
         \texttt{Hyperz}-galaxy & 0.05 & 17.8 & 17.8  \\
         \texttt{Hyperz}-AGN & 0.112 & 32.9 & 22.4   \\
         \hline \\
         \texttt{Pdz} sum & 0.049 & 16.9 & 19.2  
    \end{tabular}
    \caption{Values of $\sigma_{NMAD}$, absolute outlier fraction $\eta$ and relative outlier fraction $\eta_{rel}$ for the different codes and set of templates. The last row reports the values for the final photometric redshift solution returned by the sum of the Pdz of the different codes.}
    \label{tab:zphot_quality}
\end{table}

\section{Tests on the contamination of the radio excess sample}\label{app:contamination}
\subsection{Effect of the uncertainties on $REX$}
\begin{figure}[h!]
    \centering
    \includegraphics[width=1\linewidth]{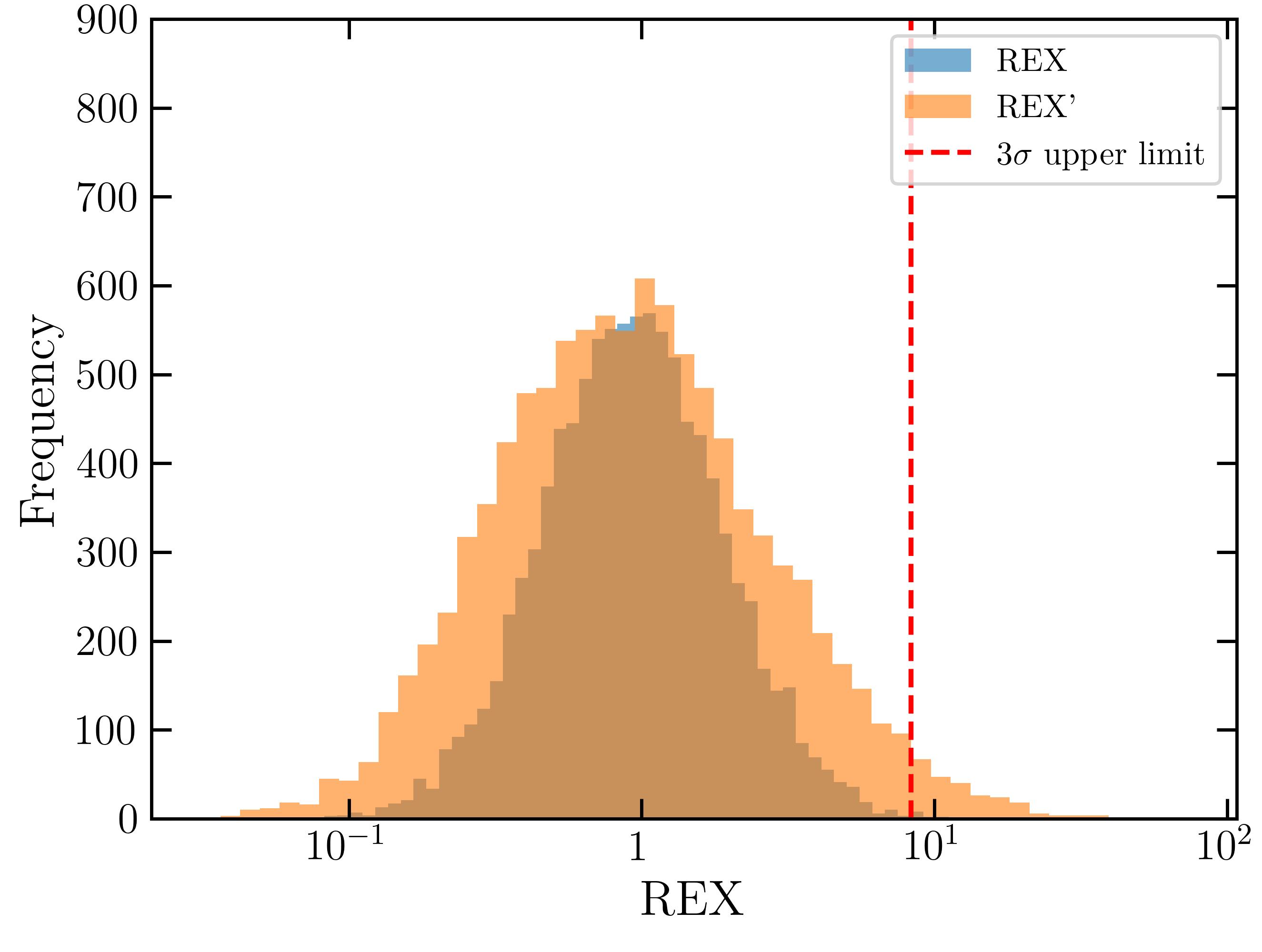}
    \caption{$REX$ (blue) and $REX$’ (orange) distribution compared to the three sigma radio excess threshold (red dashed line). REX’ values were drawn from the $REX$ distribution as described above. Sources from the $REX$' distribution exceeding the radio-excess selection threeshold do not exceed the 2.5\%.
}
    \label{fig:REX'}
\end{figure}

 There are three main quantities contributing to the uncertainties on $REX$: SFR derived from \texttt{CIGALE}, redshift, and radio flux. The first is the main source of uncertainty.
To test the impact on the radio excess selection of the uncertainties on $REX$ ($\sigma_{REX}$) we proceed as follows. 
We binned all the radio sources analyzed in the paper in $REX$ to derive the trend of $\sigma_{REX}$ with $REX$. Using error propagation, we computed $\sigma_{REX}$ from the errors on the three quantities mentioned above.
Then we performed a MonteCarlo test. We draw the Gaussian distribution of the $REX$ parameter using the same mode (peak value) derived in the paper for the SFG population and using the sigma obtained from mirroring the left part of the distribution (see Sect.~\ref{sec:radioexcessAGN}). Then, we performed 10000 random extractions of $REX$ according to the Gaussian distribution and we computed for each extracted $REX$ a new $REX$’ extracted from a second Gaussian distribution centered on the extracted $REX$ and with $\sigma=\sigma_{REX}$. We computed the fraction of $REX$’ that fulfil $REX'>8.5$, where 8.5 represents the $REX$ threshold adopted for the radio-excess selection. This fraction represents the number of sources that would be classified as radio excess just because of the effect of the error on $REX$. We found that this fraction does not exceed 2.5\% (see Fig.~\ref{fig:REX'}).
This means that from the $\sim$1000 radio sources analyzed in the paper, there can up to $\sim$20/25 sources that are real SFG and that might contaminate the radio excess selection. These might correspond up to $\sim$10\% of all the radio selected sources (that are 233) and therefore this effect is not expected to significantly affect our results. 

\subsection{Contamination from dust obscured and starburst galaxies}
The total SFR of a source is generally estimated taking into account both the direct emission from young stars as well as the contribution from obscured SF. The total SFR can be computed in two ways. The first consists of adding to the unobscured SFR observed from the rest UV-optical emission, the contribution absorbed by dust and reradiated in the IR, if photometry in the MIR and FIR is available. The second way uses the observed rest UV-optical emission, and applies the attenuation correction (always estimated from the rest-optical SED) to obtain the total SFR. This second possibility is generally applied when the MIR and FIR bands are not available.\\
The value of the $SFR_{SED}^{corr}$ we derived from the SED fitting with \texttt{CIGALE} is computed in this second way. Given that our photometry does not cover the rest-frame MIR and FIR emission of the sources, which provide the most reliable estimate of the amount of obscured SF in a galaxy, it is possible that for some sources the value of the obscuration might be underestimated. This would lead to an underestimation of $\rm SFR_{SED}^{corr}$, moving these sources to larger values of REX. 
This effect can be particularly important for the population of dust-obscured starburst galaxies, that are generally associated with a significant radio emission (related to the starburst) but whose rest UV and optical emission is almost completely absorbed by dust \citep{Talia21, Enia22, Behiri24a, Gentile24a, Gentile25}. However, the number density of these galaxies is not high enough to explain the large majority of the radio-excess selected sources on the J1030 field (Sapori et al. in prep.). \\
We can check the possible amount of dusty SFG contamination in our radio excess sample by taking advantage of the 1.1 mm AzTEC observation of the J1030 filed \citep{Zeballos18}. This observation covers an area similar to the one covered by the \textit{Chandra} image and has a median rms of 0.6 mJy. Assuming that all the radio excess sources in the \textit{Chandra} footprint are dusty SFG, we can estimate their total infrared luminosities ($L_{TIR}$) as if all their radio emission comes from SF, and therefore using the radio luminosity of Eq.~\ref{eq:Lrad} and the $q_{TIR}$ in Eq.~\ref{eq:qtir}. Taking a gray body model for the dust emission \citep[with dust temperature $T=40K$ and opacity coefficient $\beta=1.6$][]{Venemans18} normalized to the $L_{\rm TIR}$ values, we derived the expected 1.1mm flux densities of these sources. In total, we would expect 33 $>3\sigma$ detections in the AzTEC 1.1mm map of the field, while there are only 1 (4) radio-excess source whose position and expected flux density are compatible with a 1.1mm-detected source considering separations $d<3"$ ($d<6"$).\\ 
Given the depth of the AzTEC image, it is still possible that some starburst galaxies can hide among the radio excess sources without being detected in the 1.1 mm map. Therefore, we additionally performed the following test. We performed a fit with \texttt{CIGALE} inserting a module that allows a SFR burst in the last 10, 50 or 100Myr. Then, with the SFR resulting from the fit with the burst component, we repeated the procedure outlined in Sect.~\ref{sec:radioexcessAGN} and we defined a new $REX$ threshold for radio-excess sources. We ended up with 44 sources that were no longer selected as radio-excess because of the presence of a burst producing a larger (and dust-obscured) $SFR^{corr}_{SED}$. We then check the reliability of that burst component. We found that only for 29 of the 44 sources the reduced $\chi^2$ returned from the fit with the burst was lower than the one returned by the standard fit. However, a dust-obscured burst component in the fit can also be easily reproduced by a dust-obscured AGN component. Therefore, we compared the fit obtained with the burst with the fit obtained with the AGN component (see Sect.~\ref{sec:SFRdistr}), finding that for only 6 out of the 29 sources, the burst solution was preferred to the AGN one (with 4 of these having $\chi^2_{\rm AGN}-\chi^2_{\rm bust}<0.1$). These tests clearly show that the contamination from bursty SFG on our AGN selection is very small, with most of these starburst galaxies falling below the radio excess threshold, as also demonstrated by the X-ray stack in Sect.~\ref{sec:Xstack}.

\begin{figure*}
    \centering
   \includegraphics[width=0.3\linewidth]{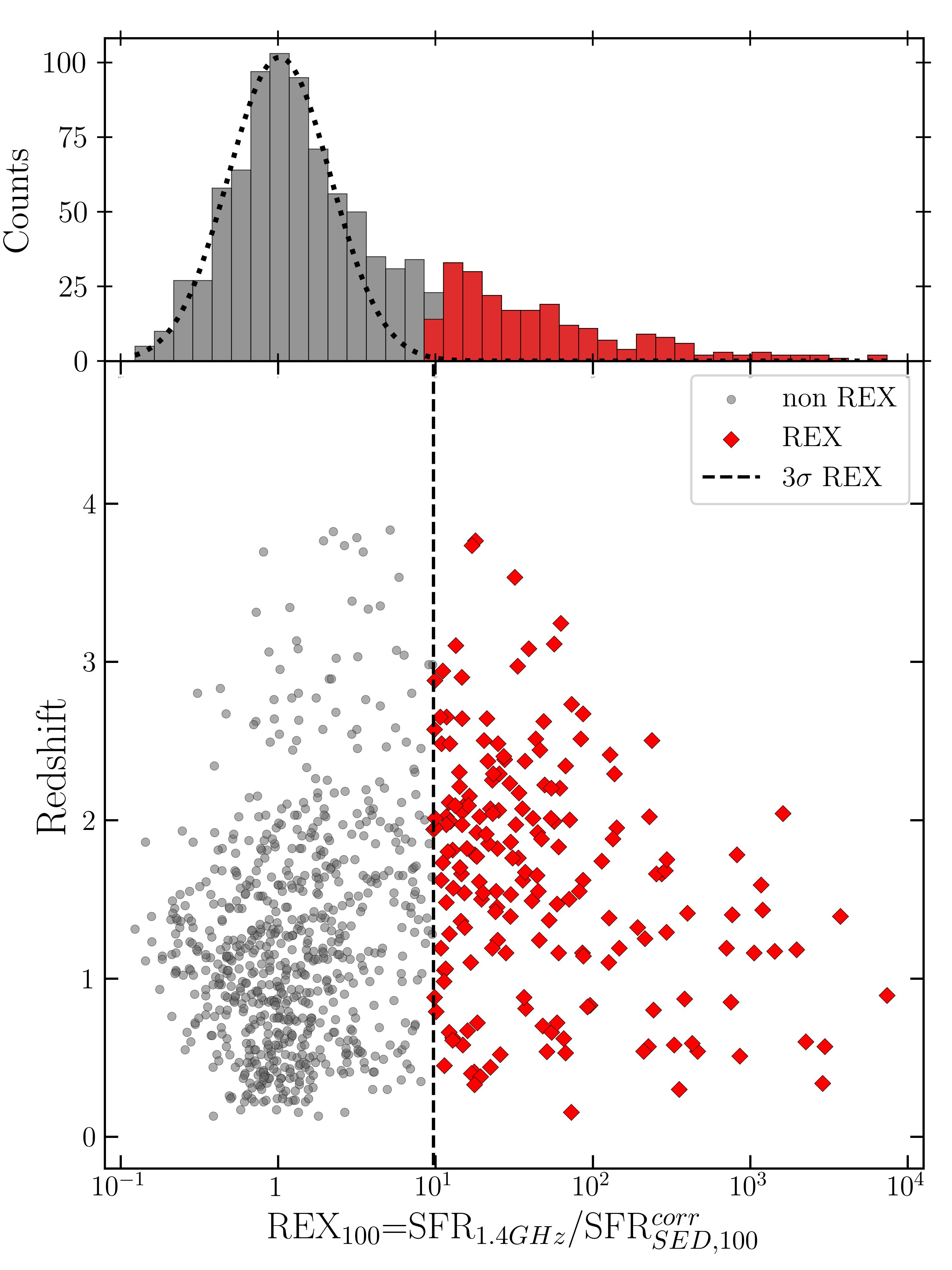}
   \includegraphics[width=0.3\linewidth]{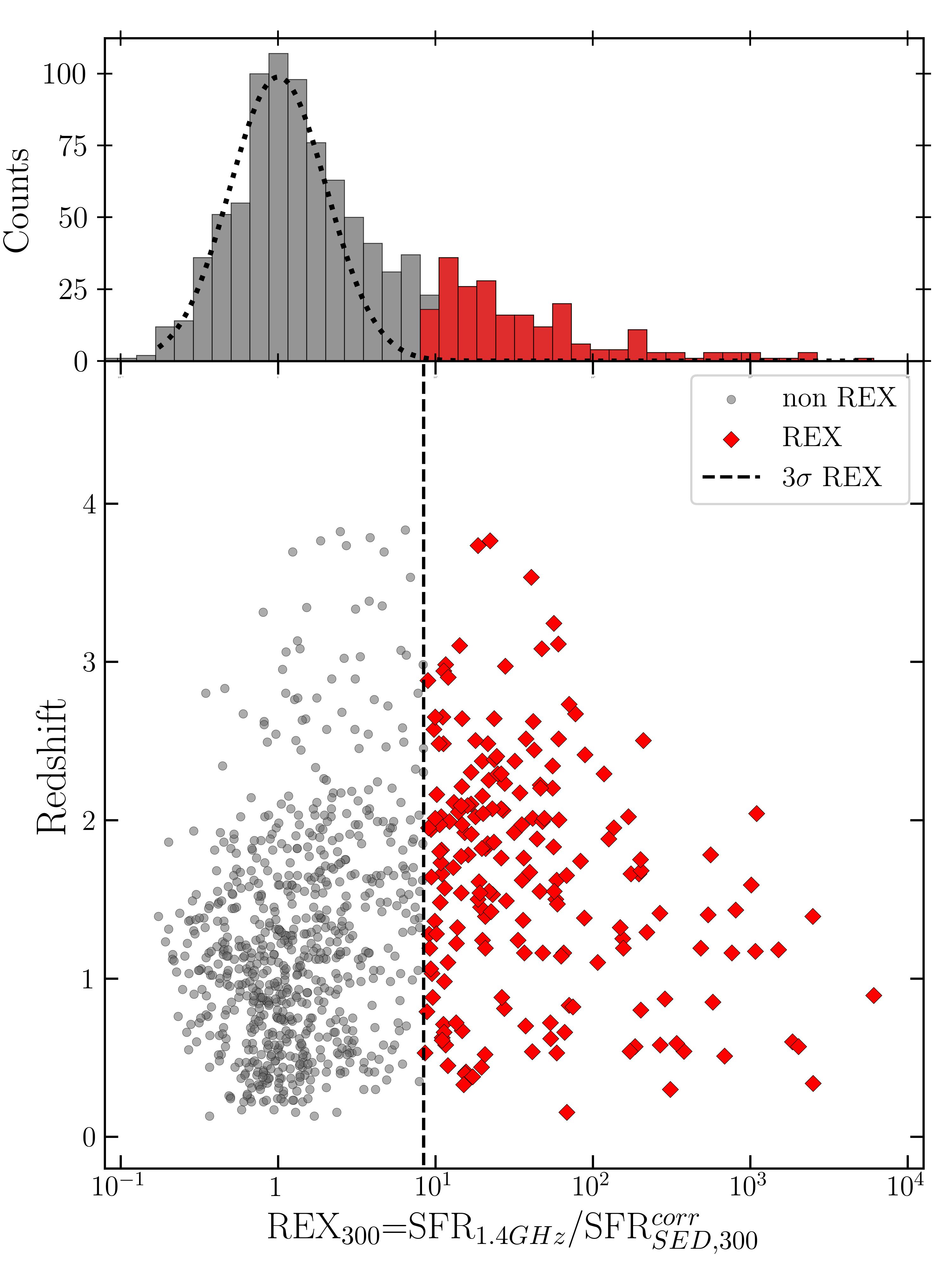}
   \includegraphics[width=0.3\linewidth]{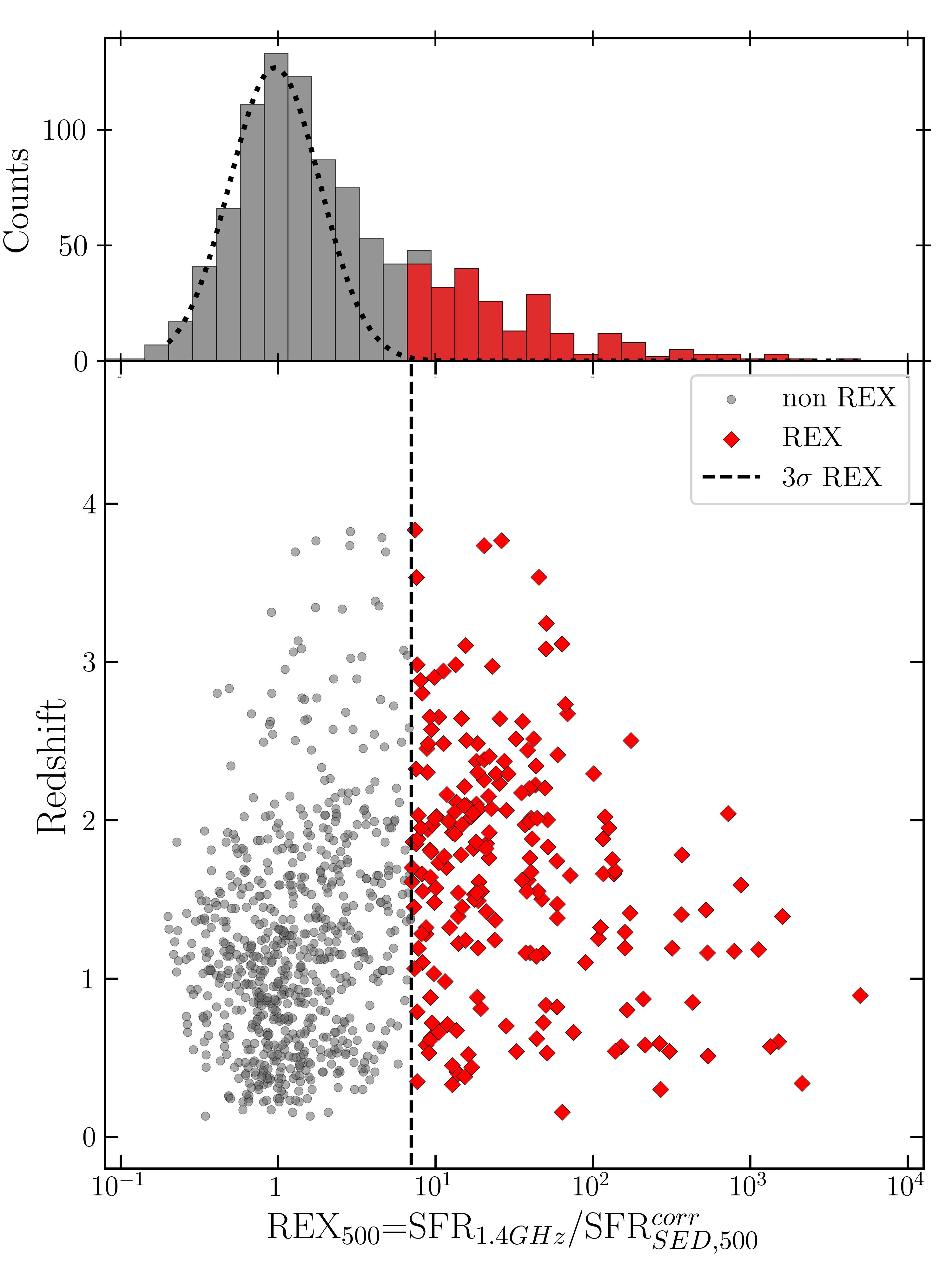}
      \caption{Same as in Fig.~\ref{fig:REXhist} but considering $SFR^{corr}_{SED}$ averaged over different timescales, as described in Appendix~\ref{app:contamination}. The peak of the distribution over $REX$ does not significantly vary in any of the three plots, as well as the radio-excess threshold is consistent for all the timescales with the one derived in Sect.~\ref{sec:radioexcessAGN}. The number of radio-excess sources is 187, 193, 209 for $REX_{100}$, $REX_{300}$, and $REX_{500}$, respectively, where the differences have to be ascribed to the small shifts of the radio-excess threshold:  $REX_{100}$=9.71, $REX_{300}$=8.44, and $REX_{500}$=7.03. }
         \label{fig:REXaverage}
\end{figure*}

\subsection{Different timescales of the radio and optical SFR}
Another possible bias of our $REX$ parameter can be the fact that the SFR traced by the radio luminosity ($SFR_{1.4 GHz}$) and the one traced by the SED-fitting ($\rm SFR_{SED}^{corr}$) refer to different timescales of star formation. In particular, \citet{ArangoToro23} showed that in some cases an excess in the radio SFR with respect to the one derived from the SED-fitting can be justified by the fact that the star formation time sensitivity of the radio frequency might be longer than 150 Myr, while the SFR derived from the optical photometry can be assumed as an instantaneous measure. Therefore sources with high values of $REX$ can be sources that might have experienced a starburst in the last $\sim 200-300$Myr followed by a decreasing SFH. Even if this possibility does not exclude that the starburst or the following quench might also be associated with AGN activity, we tested this possibility by computing with CIGALE the average SFR over the last 100, 300 and 500 Myrs, and substituting these SFR to $SFR_{SED}^{corr}$ in Eq.~\ref{eq:REX}. In Fig.~\ref{fig:REXaverage} we show the new distributions of the radio sources according to the $REX$ parameters averagred over different timescales, and as it is possible to see they are analogous to the one in Fig.~\ref{fig:REXhist}. We did not find a significant shift in the position of the sources on the x-axis or in the $REX$ threshold. \\
Another possible bias of $SFR_{SED}^{corr}$ is that the original SED-fitting with \texttt{CIGALE} was performed without including an AGN component. The unobscured AGN emission can potentially enhance the rest frame UV and optical emission, leading to an overestimation of the $SFR_{SED}^{corr}$, determining lower values of $REX$ that can shift some true AGN out from the radio-excess selection. However this is not the case for the population of heavily obscured AGN we are interested in, whose emission in the rest frame UV and optical bands is almost completely negligible.

\subsection{Contamination from LERGs}
\cite{Best23} and \cite{kondapally22,Kondapally25} showed that the population of radio excess selected AGN can be populated by radiatively inefficient radio AGN, the so-called low-excitation radio galaxies (LERGs). These sources are characterized by an inefficient accretion mechanism, producing strong radio emission and lacking the typical emission characterizing radiatively efficient AGN in the X-rays, optical, or MIR bands. Therefore the lack of X-ray emission for these sources can be an intrinsic property and not caused by obscuration. To identify this population we followed the approach outlined in \cite{Best23}, where they select as LERGs all radio excess sources without a significant radiative output in the rest frame optical or MIR bands, that is where the direct emission from the accretion disk or the reprocessed one from the dusty torus are expected to peak. Given the available photometry and accounting for energy balance \texttt{CIGALE} computes the best-fit SED over the whole EM spectrum. Considering the fit to the radio excess population performed by adding the AGN component, we defined $f_{AGN,opt}$ and $f_{AGN,IR}$ as the fractional contribution of the AGN component over the total SED (star+dust+AGN) in the $0.1-0.5 \mu m$ and in the $5-10 \mu m$, respectively. Then, to account for both obscured and unobscured AGN, we defined $f_{AGN}=max(f_{AGN,opt},f_{AGN,IR})$, and we set as a threshold for the presence of a radiatively efficient AGN $f_{AGN}=0.1$ \citep[following][]{Best23}. Our selection returned 52 radio-excess sources having $f_{AGN}<0.1$, and therefore classified as possible LERGs. While at $z<1$ $\sim 40\%$ of the radio excess sources are classified as LERGs candidates, at $z>1.5$ this fraction significantly decreases to $\sim 14 \%$ and only to the $10\%$ at $z>2.5$. The absolute number of LERGs identified by this procedure agrees well with the number of LERGs estimated on the J1030 field radio image considering the LERGs luminosity function reported in \cite{kondapally22}. However, as reported in Sect.~\ref{sec:contamination}, the LERGs selected sample includes also 7 X-ray detected sources ($15\%$ of the total LERGs sample) having $\log L_{2-10~keV}>43$ erg $s^{-1}$, that therefore clearly host radiatively efficient AGN. This implies that the actual number of LERGs should be even lower.

 \subsection{Contamination from AGN light in the SED-fitting}
Light coming from unobscured AGN might contaminate the rest-UV and optical, making the \texttt{CIGALE} SFR (estimated using only galaxy models) overestimated, because all the light is attributed only to stars. Therefore our radio-excess selection might have lost some true unobscured AGN. However, these sources are not the main target of this work. Instead, obscured AGN are not expected to significantly contribute to the rest-frame UV and optical, and therefore, the current SED-fitting should be reliable.
We tested this scenario in a more quantitative way by performing the CIGALE fit adding also the AGN component. As a result, out of the 1003 radio sources, 840 have a lower SFR (and therefore higher REX). Among the sources that we selected as radio excess and not X-ray detected, there are only 13 sources that have a REX parameter lower than our REX threshold when the AGN component is added. These are all sources for which the SED-fitting decomposition predicts a blue unobscured AGN in the rest-frame optical and UV but also a strongly dust-obscured SFR, which determines a larger CIGALE SFR and therefore a lower REX. However, given that these blue AGN are not X-ray detected, this SED-fitting decomposition is probably incorrect. At the same time, there are instead 38 sources that, with the new AGN SFR, are now in the radio excess region without being radio-excess with the previous SED-fitting decomposition and confirming that our radio-excess selection is conservative.

\section{X-ray stacking of radio sources without a Ks-band detection}\label{app:Xstack_radiound}

\begin{figure}[h!]
    \centering
    \includegraphics[width=1\linewidth]{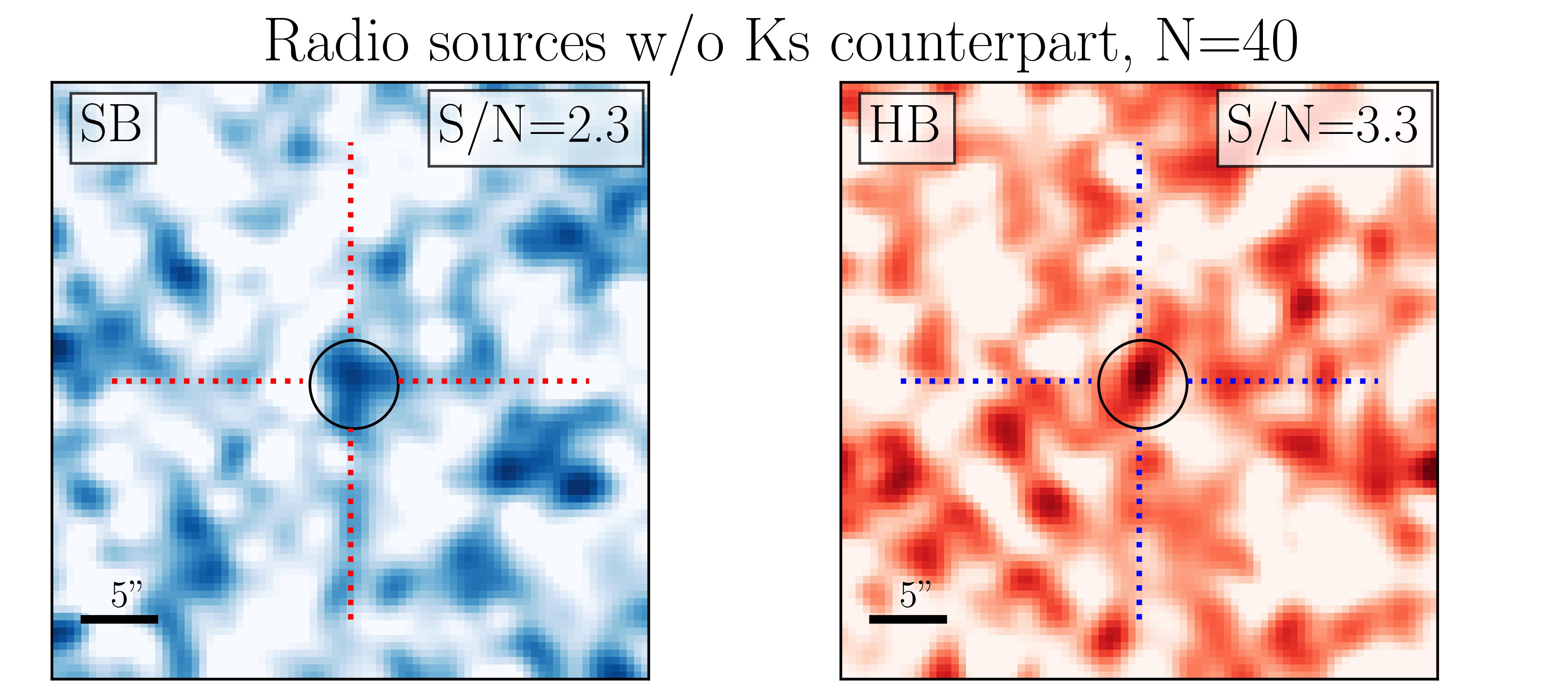}
    \caption{X-ray stacking in the 0.5-2 keV and 2-7 keV of the 40 radio detected sources with $r_{90}<6"$ and non detected in the J1030 multiband catalog.}
    \label{fig:radionly_stack}
\end{figure}
As reported in Sect.~\ref{sec:J1030rad}, there are 99 sources without a counterpart in the Ks-band selected catalog of the J1030 field. In particular, 22 of these sources are without a counterpart in any band, while most of the others are non detected (or extremely faint) in the Ks band but are visible at longer wavelengths. In any case, we could not attribute a redshift or other physical property to any of these sources, but we can still investigate the presence of obscured AGN using X-ray stacking. We again considered only sources in the \textit{Chandra} footprint (76 sources) and we stacked only those with $r_{90}<6"$ (40 sources). The results are shown in Fig.~\ref{fig:radionly_stack}. The stacking revealed a faint detection in the 2-7 keV and not in the 0.5-2 keV. Using Eq.~\ref{eq:HR}, we derived an HR$\sim 0.1\pm0.1$ that is indicative of obscuration, even if for a more robust estimate, redshifts are needed. We also performed an X-ray stacking analysis considering only the radio sources without a counterpart in any band (in total 22 sources 18 in the X-ray footprint). However, the X-ray stacking did not reveal any detection for these sources, probably due to the limited statistics or the very high obscuration. \\
In any case, the X-ray stacking showed that at least a fraction of these Ks-band non detected radio sources can contribute to the most obscured AGN population missed by the X-ray observation but revealed by deep radio continuum imaging, further contributing to the population of radio detected heavily obscured AGN.

\end{appendix}

\end{document}